\documentclass[11pt]{article}
\usepackage{jcappub,natbib,float,caption}

\usepackage{graphicx,amsmath,amssymb,color,bm}
\usepackage[dvipsnames]{xcolor}
\usepackage[caption=false]{subfig}
\usepackage{arydshln,verbatim}
\usepackage{rotating}

\usepackage{tikz}
\usetikzlibrary{arrows,intersections,patterns,decorations.markings,shapes.geometric}
\usetikzlibrary{calc,fadings,decorations.pathreplacing,positioning}

\tikzset{
  >=latex, 
  inner sep=0pt,
  outer sep=2pt,
  mark coordinate/.style={inner sep=0pt,outer sep=0pt,minimum size=3pt,
    fill=black,circle}
}

\usepackage{pgfplots}
\usepgfplotslibrary{colormaps}

\def\be{\begin{equation}}
\def\ee{\end{equation}}
\def\bea{\begin{eqnarray}}
\def\eea{\end{eqnarray}}
\newcommand{\vs}{\nonumber\\}
\def\bg#1\eg{\begin{gather}#1\end{gather}}

\newcommand{\refeq}[1]{Eq.~(\ref{eq:#1})}          
\newcommand{\refeqs}[2]{Eqs.~(\ref{eq:#1})--(\ref{eq:#2})}

\renewcommand{\v}[1]{\bm{#1}}
\def\P{\mathcal{P}}

\DeclareMathOperator{\tr}{tr}
\DeclareMathOperator{\TF}{TF}

\newcommand{\vx}{\v{x}}

\newcommand{\vr}{\v{r}}
\newcommand{\vk}{\v{k}}

\newcommand{\<}{\langle}
\renewcommand{\>}{\rangle}

\renewcommand{\d}{\delta}
\def\dK{\delta^\text{K}}
\def\dD{\delta^\text{D}}

\newcommand{\vhat}[1]{\v{\hat{#1}}}

\newcommand{\Om}{\Omega_m}

\def\Del{\mathcal{D}}

\def\cH{\mathcal{H}}

\renewcommand{\vec}{\bm}
\def\vr{\vec{r}}
\newcommand*{\p}  {\partial}
\newcommand*{\pp}  {\parallel}
\newcommand*{\df}  {\delta}

\newcommand*{\non}  {\nonumber}
\newcommand*{\lb}  {\left(}
\newcommand*{\rb}  {\right)}

\newcommand*{\la}  {\left\langle}
\newcommand*{\ra}  {\right\rangle}

\newcommand{\ba}{\[\begin{aligned}}
\newcommand{\ea}{\end{aligned}\]}

\newcommand{\eq}[1]{\begin{align}#1\end{align}}
\newcommand{\eeq}[1]{\begin{equation}#1\end{equation}}

\title{Galaxy shape statistics in the effective field theory}

\author{Zvonimir Vlah$^{[a,b,c]}$,}
\author{Nora Elisa Chisari$^{[d]}$,}
\author{Fabian Schmidt$^{[e]}$}

\affiliation[a]{Kavli Institute for Cosmology, University of Cambridge, Cambridge CB3 0HA, UK.}
\affiliation[b]{Department of Applied Mathematics and Theoretical Physics, University of Cambridge, Cambridge CB3 0WA, UK.}
\affiliation[c]{Theory Department, CERN, 1 Esplanade des Particules, CH-1211 Gen\` eve 23, Switzerland.}
\affiliation[d]{Institute for Theoretical Physics, Utrecht University, Princetonplein 5, 3584 CC Utrecht, The Netherlands.}
\affiliation[e]{Max-Planck-Institut f\"ur Astrophysik, Karl-Schwarzschild-Str. 1, 85741 Garching, Germany.}

\emailAdd{zv217@cam.ac.uk}
\emailAdd{n.e.chisari@uu.nl}
\emailAdd{fabians@mpa-garching.mpg.de}

\abstract{  
  Intrinsic galaxy alignments yield an important contribution to the observed statistics of galaxy shapes. The general bias expansion for galaxy sizes and shapes in three dimensions has been recently described by Vlah, Chisari \& Schmidt using the general perturbative effective field theory (EFT) framework, in analogy to the clustering of galaxies. In this work, we present a formalism that uses the properties of spherical tensors to project galaxy shapes onto the observed sky in the flat-sky approximation, and compute the two-point functions at next-to-leading order as well as the leading-order three-point functions of galaxy shapes and number counts. The resulting expressions are given in forms that are convenient for efficient numerical implementation. For a source redshift distribution typical of Stage IV surveys, we find that that nonlinear intrinsic alignment contributions to galaxy shape correlations become relevant at angular wavenumbers $l \gtrsim 100$.
}

\begin{document}

\maketitle
\flushbottom

\section{Introduction}
\label{sec:intro}

The advent of wide and fast galaxy surveys has triggered an era of precision cosmology from the large-scale structure, 
with this cosmological tool being used to elucidate the nature of the dark components of the Universe and the physics 
of the early Universe, among other goals. But in order for the inferences drawn from ever more powerful data sets to be robust, 
a similar progression must occur on the modelling side: our model for the observables needs to be accurate enough 
to guarantee no residual biases in the analyses.

Weak gravitational lensing is one of the phenomena observed in the large-scale structure which can yield information 
about the nature of dark matter and the origin of the accelerated expansion of the Universe. Gravitational lensing by 
the large-scale structure of the Universe distorts the shapes of background sources (galaxies) in a correlated manner. 
Although the signal is challenging to extract (amounting to percent-level correlated changes in galaxy ellipticities), it is now 
regularly and successfully extracted from multiple data sets, and a number of collaborations have presented cosmological
constraints from this probe in the last decade 
\cite{Heymans++:2013, Jee++:2012, Huff++:2011, Hildebrandt++:2016, Troxel++:2017, Hildebrandt++:2018, Joudaki++:2019}. 

One of the challenges in the modelling of galaxy shape correlations is the coherent distortion induced by gravitational 
interactions between galaxies and the large-scale distribution of matter in the Universe. Such physical distortions are 
known as ``intrinsic alignments'' and have been detected to high significance by multiple experiments, with an increasing 
body of work focusing on which, how and when galaxies align, and how these alignments give rise to a contamination 
of the weak lensing signal. On the other hand, their detection is on such firm grounds 
\cite{Mandelbaum++:2005,Hirata++:2007,Okumura++:2008,Singh++:2014} 
that there are proposals for their use as a cosmological probe on their own right 
\cite{Chisari++:2014, Schmidt++:2015, Chisari++:2016b, Akitsu++:2020}. 

So far, the modelling of intrinsic galaxy alignments has lagged behind similar efforts for other large-scale structure probes. 
The most commonly adopted model is the ``linear alignment model'' (LA) \cite{Catelan++:2000}, which postulates a linear 
relation between the projected shapes of galaxies and the projected tidal field of the large-scale structure. This
ansatz in any case provides a good description at large separations only. In the 
nonlinear regime, an excess above LA predictions is observed for early-type galaxies, which can be well-approximated by an {\it ad hoc} replacement 
of the linear matter power spectrum by its nonlinear analogue, while preserving the assumption of linear alignment \cite{Bridle++:2007}. 
An alternative for nonlinear modelling is the so-called ``halo model'', by which the small-scale regime is modelled through 
the distribution of aligned central and satellite galaxies within a halo \cite{Schneider++:2009, Joachimi++:2012, Joachimi++:2013}. On the other 
hand, disk galaxies are assumed to be subject to alignments via tidal torques, and a separate set of models has been 
constructed to describe this flavor of alignments \cite{Heavens++:2000, Crittenden++:2000, Mackey++:2001}. Cosmological 
simulations serve as a validation tool for all these models and help constrain their free parameters in as much as the simulated 
galaxy population is representative of the observed one 
\cite{Kiessling++:2015, Codis++:2014, Chisari++:2015, Chisari++:2016a, Velliscig++:2015a, Velliscig++:2015b, Tenneti++:2014a, 
Tenneti++:2014b, Tenneti++:2015a, Tenneti++:2016, Chisari++:2017, Hilbert++:2016, Taruya+:2020, Kurita++:2020}.

The increasing precision of galaxy shape measurements requires, however, new modelling tools if we are to ensure the 
success of weak lensing cosmology \cite{Krause++:2015}. Recently, a new theory approach was put forward that unifies 
the two alignment mechanisms under the same framework by means of standard perturbation theory (SPT) 
\cite{Blazek++:2015, Blazek++:2017,Schmitz++:2018}. 
This more rigorous approach expands over the existing models and has already been applied to observational 
data sets \cite{Troxel++:2017, Samuroff++:2018}. In \cite{Vlah++:2019}, we presented a complete formulation that is based 
on a similar expansion in the context of the ``effective field theory'' (EFT) of the large-scale structure. Considering the symmetries 
of a trace-free tensor, we were able to identify all potential gravitational observables at a given order that contribute to describing 
any given biased tensorial tracer of the large-scale structure, including the intrinsic shapes of galaxies. 

The EFT framework we presented in \cite{Vlah++:2019} focused on three-dimensional intrinsic shapes and their correlations. 
In this work, we connect the three-dimensional observables computed in that work to the projected ones from galaxy surveys, 
namely: quantities corresponding to the convergence field $\kappa$ and shear fields $\gamma_1$ and $\gamma_2$. 
We present results for the one-loop angular power spectrum and tree-level angular bispectrum for the convergence and 
shear auto- and cross-correlations. We emphasise, however, that the formalism derived in this work does not rely explicitly 
on the EFT expansion. Our findings can be applied to projecting the statistics of any three-dimensional scalar and tensorial 
fields regardless of how one chooses to model them. In this paper, we do not include nonlinear projection contributions, 
such as those from reduced shear. We will argue below that these are much smaller than the nonlinearities in intrinsic 
shape correlations which we treat here.

This work is organized as follows. In Section \ref{sec:bias}, we summarise the formalism presented in Paper I \cite{Vlah++:2019}. 
We define the projected observables in Section \ref{sec:proj}. We construct predictions for intrinsic alignment two-point and three-point 
correlations assuming the flat-sky approximation in Section \ref{sec:flat-sky}. 
We produce numerical predictions for the typical observables and redshifts of galaxy surveys in Section \ref{sec:res}. 
Conclusions are presented in Section \ref{sec:concl}.  
The contribution of gravitational lensing to the two-point shape statistics are included assuming the flat-sky approximation in Appendix \ref{app:lensing}.
Details of the projection calculation for the angular power spectrum are presented in Appendix \ref{app:flat_sky_spectrum}
and similarly for the angular bispectrum in  Appendix \ref{app:flat_sky_bispectrum}.
  
Throughout the paper, we assume an Euclidean $\Lambda$CDM cosmology with $\Omega_m=0.272,~\Omega_b=0.0455,~n_s=0.967,
~\sigma_8=0.807$ and $h=0.704$.  We work under the assumptions of adiabatic Gaussian perturbations and General Relativity. 
It is straightforward to include the impact of primordial non-Gaussianity following previous work \cite{Schmidt++:2015}. 
In perturbative calculations, we always make the usual approximation of setting the $n$-th order growth factor $D^{(n)}(\tau)$ to $[D(\tau)]^n$.

\begin{table*}
\centering
\begin{tabular}{l|l}
\hline
\hline
$\dK_{ij}$ & Kronecker symbol \\[3pt]
$\dD(\vx)$ & Dirac delta function \\
$\Del_{ij} \equiv \p_i\p_j/\p^2 - (1/3)\dK_{ij}  $ & Shear derivative \\
$\vk_{1\cdots n} \equiv \vk_1 + \cdots + \vk_n$ & Sum notation\\
$\vec Y_{ij}^{(m)}(\vhat k)$ & Rank-two spherical tensor basis in Fourier space\\
$\vec M_{ij}^{(m)}(\vhat r)$ & Rank-two spherical tensor basis in configuration space\\
$f(\vk) \equiv \int d^3 \vx\, f(\vx) e^{-i\vk\cdot\vx}$  & Fourier transform \\
$\< O(\vk_1) \cdots O(\vk_n)\>'\,$
 & $n$-point correlator without momentum conservation\footnotemark\\[2pt]
\hline
$\d_{\rm m}$ & Fractional matter density perturbation \\
$\Pi^{[1]}_{ij} \equiv \Del_{ij} \d_{\rm m} + (1/3)\dK_{ij} \d_{\rm m}$ &
Scaled Hessian of gravitational potential  \\
$v^i$ & Matter velocity field \\
\hline
$\d_{\rm n}$ & Fractional galaxy number density perturbation \\
$\d_{\rm s}$ & Fractional galaxy size density perturbation \\
& \qquad (trace of shape tensor) \\
$S_{ij}$ & 3D galaxy shape tensor \\
$g_{ij}$ & Trace-free part of 3D galaxy shape tensor \\
$\gamma_{ij}$ & Projected shape on the sky \\
& \qquad (trace-free part of projected shape tensor) \\
$\vhat r$ & Unit vector along the line of sight \\
$\vhat n$ & Fixed unit vector perpendicular to the flat sky plane \\
\hline
\end{tabular}
\caption{List of notation and most important quantities used in this paper.
  Fields in Fourier space are understood to be integrated over repeated
  momentum variables.
} 
\label{tab:notation}
\end{table*}
\footnotetext{Explicitly, we write 
\eeq{ \la O(\vk_1) \cdots O(\vk_n)\ra = (2\pi)^3 \d^D(\vk_{1\cdots n}) \la O(\vk_1) \cdots O(\vk_n)\ra' \,,}
where the Dirac delta function ensures the total momentum conservation, and is a consequence of the statistical translation invariance.}

\section{EFT of intrinsic galaxy shapes}
  \label{sec:bias}
  
\subsection{Galaxy shapes and bias expansion}
\label{sec:galaxy_shapes}

The three-dimensional shapes of galaxies in their rest frame are spin-$2$ quantities that can be described in terms of a 
well-defined expansion in local gravitational observables. In Paper I \cite{Vlah++:2019}, we assumed that, for each galaxy $\alpha$, 
its light distribution is described by the symmetric second-moment tensor $I_{ij}(\vr_\alpha)$ (i.e., an ellipsoid) and defined by
the shape fluctuation field\footnote{Different normalizations of the $S_{ij}$ field are possible, though this does not affect the form or range of validity of the EFT expansion.}
\eeq{
S_{ij}(\vr) = g_{ij}(\vr) + \frac13 \d_{\rm s}(\vr) \dK_{ij}\,,
\label{eq:Sij}
}
where the trace-free tensor $g_{ij}$ describes galaxy shape perturbations, and the scalar field $\d_{\rm s}$, galaxy size perturbations. 
We presented a new EFT expansion for the former in Paper I \cite{Vlah++:2019}, while the latter is described by the usual EFT of 
scalar biased tracers \cite{Mirbabayi++:2014, Desjacques++:2016} in complete analogy to the case of galaxy number counts. This formalism allowed us to provide 
expressions for the two-point correlations between the intrinsic galaxy shape field and other scalar fields up to next-to-leading order, 
for which we drew analogies with the EFT application to scalar quantities, such as the number density of biased tracers \cite{Mirbabayi++:2014, Desjacques++:2016}. 
Moreover, we presented three-point functions for these tracers at leading order.
The EFT approach also shows that one has to allow for distinct bias parameters in the bias expansion for the trace (size) and the trace-free (shape) parts.

Explicitly, we expand the number counts $\delta_{\rm n}$, size fields $\delta_{\rm s}$,
and shape fields $g_{ij}$ in terms of local gravitational observables as follows
\eq{
\label{eq:d_and_gij}
\hspace{-2cm} a \in \{ {\rm n},\, {\rm s} \}: ~~  \d_{a}(\vk,\tau) &= \sum_O b_O^{(a)}(\tau) \tr [O_{ij}](\vk,\tau)\,, \\
  g_{ij}(\vk,\tau) &= \sum_O b_O^{({\rm g})}(\tau) \TF [O_{ij}](\vk,\tau)\,, \non
}
where $\tr$ and $\TF$ denote the trace and trace-free components of a tensor, respectively, and $[O]$ are renormalized operators while $b_O^{(a)}$ are the corresponding (renormalized) bias parameters. 
The expansion is not unique, but if the physical processes that determine the properties of the tracers are {\it local}, 
reference \cite{Mirbabayi++:2014} showed that one possible complete basis is the one comprised of all scalar 
combinations of a set of operators $\Pi^{[n]}$, defined recursively starting from 
\eeq{
\Pi^{[1]}_{ij}(\vr,\tau) = \frac{2}{3\Om\cH^2} \partial_{x,i}\partial_{x,j}\Phi(\vr,\tau)
= \frac13 \dK_{ij}\, \d_{\rm m} +\Del_{ij} \d_{\rm m}\,.
\label{eq:hatPi}
}
Note that Eq.~(\ref{eq:hatPi}) contains the leading gravitational observables at a given spacetime position $\vr,\tau$: namely, the matter 
density perturbation $\d_{\rm m}$, and the scaled tidal field $\Del_{ij} \d_{\rm m}$. In the case of shapes, the expansion must account for all possible 
trace-free tensor combinations, which in general results in more contributions at a given order than in the scalar (density) case. 

The computation of correlations between biased tensorial and scalar fields can be simplified by considerations of isotropy. To this end, 
in Paper I \cite{Vlah++:2019} we proposed to decompose any given tensorial field in spherical tensors \cite{Sakurai+:2011} of multipole 
order $\ell=2$, whose transformation properties under rotation are known.
We thus decompose the shape tensor field into these basis functions in terms of helicity $m=0,\pm1,\pm2$ defined with respect to the wavevector $\vk$:
\eq{
\label{eq:SVT_decomp}
S_{ij} (\vec k) 
&= \sum_{\ell = 0,2} \sum_{m=-\ell}^\ell S^{(m)}_\ell (\vec k) \left(\vec{Y}^{(m)}_\ell(\vhat k)\right)_{ij} \\
&= \frac{1}{3}S^{(0)}_0 (\vec k) \dK_{ij} Y^{(0)}  + \sum_{m=-2}^2S^{(m)}_2 (\vec k) \vec Y^{(m)}_{ij}(\vhat k), \non
}
where $\vec Y^{(0)}_{0, ij} = Y^{(0)} \dK_{ij}$, and $Y^{(0)}=1$ is the single scalar helicity-0 mode (explicitly written in this form for the purpose of symmetry). 
Hereafter, we will drop the tensor indices $\ell$ on the basis vectors and use the boldface symbol $\vec{Y}$ in order to emphasize that we are dealing with 
a basis  $\ell=2$ tensor, while for the $\ell=0$ component we will use the explicit form given in Eq. (\ref{eq:d_and_gij}). Note that the spherical tensor components $S^{(0)}_0 (\vec k)$ 
and $S^{(m)}_2 (\vec k)$ can be directly obtained from $S_{ij}$ by projections using the basis tensors.

To construct the basis explicitly, we start by defining an orthonormal basis of three-dimensional Euclidean space comprised of tensors
\eeq{
\vec e_1= \frac{\vec k \times \v{\hat{n}}}{|\vec k \times \v{\hat{n}} |},~~ \vec e_2 = \vhat k \times \vec e_1~~ \vec e_k = \vhat k\, ,
\label{eq:k_basis}
}
where $\v{\hat{n}}$ can be chosen as the line of sight direction.
With this basis defined, we are able to build our helicity basis,
\be
\vec e^0 = \vec e_k,~~
\vec e^\pm = \mp \tfrac{1}{\sqrt 2} \lb \vec e_1 \mp i \vec e_2 \rb\,.
\ee
As a consequence, the orthonormal tensor basis functions can be defined as
\eq{
&\vec Y^{(0)}_{ij} = N_0 \Big( \hat k_i \hat k_j - \tfrac{1}{3} \dK_{ij} \Big), 
\qquad \vec Y^{(\pm 1)}_{ij} = N_1 \Big( \hat k_j e_i^{\pm} + \hat k_i e_j^{\pm} \Big),
\qquad \vec Y^{(\pm 2)}_{ij} = N_2 e_i^{\pm} e_j^{\pm},
\label{eq:Y_basis}
}
where the normalization is chosen as $N_{0,1,2}=\left\{ \sqrt{\tfrac{3}{2}}, \sqrt{\frac{1}{2}}, 1\right\}$.
$\vec{Y}^{(m)}_{ij}$ are trace-free, and complex conjugation gives 
$\vec Y^{(m)*}_{ij} = (-1)^{m}  \vec Y^{(-m)}_{ij}$, while a similar relation holds for the basis $\{ \vec e_\pm, \vec e_0 \}$.

\subsection{Two-point functions in 3D}

In Paper I \cite{Vlah++:2019}, we considered the auto-correlations of scalar and trace-free tensor fields, and their cross-correlation, defined as
\eq{
\label{eq:contributions_ss_sg_gg}
  \< \d_a(\vk) \d_b(\vk') \>' &= (2\pi)^3 \dD_{\vec k+\vec k'}P^{ab}(k)\,,\quad\mbox{where}\quad
  a,\,b \in \{{\rm n},\, {\rm s} \} \\
\< \d_a(\vk) g_{ij}(\vk') \>' &= (2\pi)^3 \dD_{\vec k+\vec k'}P^{a{\rm g}}_{ij}(k) \vs
\< g_{ij}(\vk) g_{kl}(\vk') \>' &= (2\pi)^3 \dD_{\vec k+\vec k'}P^{\rm gg}_{ijkl}(k) \,. \non
}
and we obtained explicit expressions for these by using the EFT expansion of the shape field (Eq. \ref{eq:d_and_gij}) and projecting out the 
trace and trace-free part of the power spectra. To simplify the calculation, we made use of the spherical tensor basis given in 
\refeq{SVT_decomp} and obtained these power spectra by combination of the ones corresponding to the spherical tensor field 
components $S^{(m)}_2$, which are invariant quantities that also transform as spherical tensors. 
For projecting onto the flat sky (Section \ref{sec:flat-sky}), we can directly operate with the power spectra of the $S^{(m)}_2$ components, 
which we proceed to summarize in what follows.

Due to statistical isotropy and homogeneity, the component power spectra are given by
\eeq{
\la S_\ell^{(m)}(\vec k) ~S_{\ell'}^{(m')} (\vec k') \ra = (2\pi)^3 \dK_{mm'} \dD_{\vec k + \vec k'} P^{(m)}_{\ell \ell'} (k),
\label{eq:PS_SS}
}
where $\ell,\ell'=0,2$ represent the trace and trace-free components, respectively, and $m,m'$ are helicity components. Different helicity 
components do not correlate. This is a consequence of statistical isotropy. Because the shape tensor field is real, and we 
assume parity invariance, this implies that the power spectra of the helicity components must satisfy: $P^{(m)}_{\ell \ell'} (k) = P^{(-m)}_{\ell \ell'} (k)$. 
As a consequence, the shape-shape tensor  power spectrum can be decomposed into six independent contributions, 
\eq{
P^{SS}_{ijkl}(\vec k)
= \frac{1}{9} \dK_{ij} \dK_{kl}  P^{(0)}_{00}(k)
+   \frac{1}{3}  \dK_{\{ij} \vec Y^{(0)}_{kl\}} P^{(0)}_{02}(k)
+ \vec Y^{(0)}_{ij} \vec Y^{(0)}_{kl}  P^{(0)}_{22}(k) + \sum_{q=1}^2 (-1)^q \vec Y^{(q)}_{\{ij} \vec Y^{(-q)}_{kl\}} P^{(q)}_{22}(k), 
\label{eq:decomp_ps_main}
}
where the last term is symmetrized over $(q)$ and $(-q)$ contributions,
and the curly brackets in the subscripts denote symmetrization over the two pairs of indices.

We are now able to take the trace or trace-free components to obtain expressions for the power spectra defined in Eq. \eqref{eq:contributions_ss_sg_gg}, for which we obtain
\eq{
\label{eq:spectra_all}
P^{ab}(k) & = P^{ab(0)}_{00} (k),\quad\mbox{where}\quad  a,\,b \in \{ {\rm n},\, {\rm s} \}  \\
P^{a{\rm g}}_{ij} (\vec k) & = \vec Y^{(0)}_{ij}(\vhat k) P^{a{\rm g}(0)}_{02}(k), \non\\
P^{\rm gg}_{ijkl} (\vec k) & = \vec Y^{(0)}_{ij}(\vhat k) \vec Y^{(0)}_{kl}(\vhat k) P^{{\rm gg}(0)}_{22}(k) 
+ \sum_{q=1}^2 (-1)^q  \vec Y^{(q)}_{\{ij}(\vhat k) \vec Y^{(-q)}_{kl\}}(\vhat k) P^{{\rm gg}(q)}_{22}(k)\,. \non
}
Explicit expressions for the spherical component power spectra up to one-loop order in perturbation theory were provided 
in Section 5 of Paper I \cite{Vlah++:2019} relying on the EFT expansion. Nevertheless, the decomposition in 
spherical tensors that we presented in Paper I \cite{Vlah++:2019} and summarized here is valid nonlinearly for any order in PT.

\subsection{Three-point functions in 3D}
\label{sec:three-point functions_in_3D}

We now consider three tensor fields, each with a trace and trace-free part. The bispectrum of these fields is defined by
\eq{
\la S_{ij} (\vec k_1)  S_{kl} (\vec k_2) S_{rs} (\vec k_3) \ra = (2\pi)^3 \df^{\rm D}_{\vec k_1 + \vec k_2 + \vec k_3} B^{SSS}_{ijklrs} (\vec k_1, \vec k_2, \vec k_3),
\label{eq:Bispectrum_3D}
}
Given the decomposition of $S_{ij}$ into spherical tensor components (Eq.~\eqref{eq:SVT_decomp}), we can also define the bispectrum of a combination of the components as
\eq{
\la S_{\ell_1}^{(m_1)}(\vec k_1) ~S_{\ell_2}^{(m_2)}(\vec k_2)~S_{\ell_3}^{(m_3)}(\vec k_3) \ra = (2\pi)^3 \df^{\rm D}_{\vec k_1 + \vec k_2 + \vec k_3} B^{(m_1 m_2 m_3)}_{\ell_1 \ell_2 \ell_3} (\vec k_1, \vec k_2, \vec k_3).
\label{eq:Bellellellm}
}
As a result, the decomposition of the bispectrum in terms of the $\vec Y^{(m)}$ basis tensors is then
\eq{
B^{SSS}_{ijklrs} (\vec k_1, \vec k_2, \vec k_3)
&= \frac{1}{27} \dK_{ij} \dK_{kl}\dK_{rs}  B^{(0,0,0)}_{000} (\vec k_1, \vec k_2, \vec k_3) \label{eq:bis_decomp_0}  \\
&~~~ +   \frac{1}{9}  \dK_{ij}  \dK_{kl} \sum_{m_3=-2}^2 \vec Y^{(m_3)}_{rs}\big( \vhat  k_3 \big)  B^{(0,0,m_3)}_{002} (\vec k_1, \vec k_2, \vec k_3)
 +  {\rm 2~cyc. }\non\\
&~~~ +   \frac{1}{3}  \dK_{ij} \sum_{\substack{m_i=-2 \\ i=(2,3)}}^2 \vec Y^{(m_2)}_{kl}\big( \vhat  k_2 \big) \vec Y^{(m_3)}_{rs}\big( \vhat  k_3 \big) 
B^{(0,m_2,m_3)}_{0 2 2} (\vec k_1, \vec k_2, \vec k_3)
 +  {\rm 2~cyc. } \non\\
&~~~ + \sum_{\substack{m_i=-2 \\ i=(1,2,3)}}^2 \vec Y^{(m_1)}_{ij}\big( \vhat  k_1 \big) \vec Y^{(m_2)}_{kl}\big( \vhat  k_2 \big) \vec Y^{(m_3)}_{rs} \big( \vhat  k_3 \big) 
 B^{(m_1, m_2, m_3)}_{222} (\vec k_1, \vec k_2, \vec k_3), \non
}
with 2 cyclic permutations as indicated. 
Statistical rotation and parity invariance give additional requirements that these bispectra need to satisfy, namely
\eeq{
\label{eq:bis_parity}
B^{(m_1,m_2,m_3)}_{\ell_1 \ell_2 \ell_3} = (-1)^{m_1+m_2+m_3} B^{(-m_1,-m_2,-m_3)}_{\ell_1 \ell_2 \ell_3}.
}
This significantly reduces the number of the independent bispectra of spherical tensor components in the expansion 
of Eq.~(\ref{eq:bis_decomp_0}).

The EFT framework developed in Paper I \cite{Vlah++:2019} allowed us to find all the independent 
contributions to each bispectrum $B^{(m_1,m_2,m_3)}_{\ell_1 \ell_2 \ell_3}$. We refer the reader to that work 
for the explicit expansion of the three-dimensional bispectra of galaxy densities, sizes and shapes. 
Here, we will show how these bispectra, generically Eq.~(\ref{eq:Bispectrum_3D}), can be projected 
on the sky with the help of the spherical tensor decomposition of Eq.~(\ref{eq:bis_decomp_0}).

The observables we are interested in are auto- and cross-correlators of galaxy densities, sizes and shapes. 
As mentioned in Paper I \cite{Vlah++:2019}, one should note that helicity $\pm 1$ and $\pm 2$ 
can in principle contribute to any of the bispectra where at least one shape field is correlated. 
This is due to symmetry constraints being less stringent in the bispectrum, when compared to the power spectrum.

\section{Statistics of shape fields projected on the sky}
\label{sec:proj}

In imaging surveys we usually do not have access to the 3D shape of galaxies that we have thus far been describing.
We can measure only 2D images, obtained by projections of the 3D shapes on the sky plane. Thus, in order to connect 
our theoretical models with the measured shape $\gamma_{ij}$ these sky-projections need to be taken into account.  
If we introduce the projector
\eeq{
\mathcal P_{ij}(\vhat r) \equiv \dK_{ij} - \hat r_i \hat r_j ,
} 
where $\vhat r$ is the direction pointing along the line of sight on the sky,
we can write the total shape field on the sky as
\eeq{
\gamma_{ij}(\vr^s,z) = \TF\left[\P_{ik}\P_{jl} ~ g_{kl}(\vr[\vr^s],z)\right]
+ \gamma_{G,ij}(\vr[\vr^s],z)\,,
\label{eq:g_projected}
}
where $\gamma_{G,ij}$ denotes the weak lensing shear, and $\vr^s$ and $\vr$ are respectively the 
redshift space and real space coordinates.\footnote{Note that the projectors $\P_{ij}$ are invariant 
on the redshift space mapping, i.e. 
$$ \P_{ik}(\vhat r) = \P_{ik} (\vhat r^s), $$
since the directions of real and redshift space coordinates coincide.} 
At linear order lensing shear can be simply added to the intrinsic shape, however this holds only at leading order. 
In general, the observed shape field is a nonlinear function of shear and intrinsic shape; we will discuss this issue below.
We will not consider these nonlinear effects, but defer them to future work. 

Considering thus only the projections of the intrinsic shape, and neglecting redshift-space distortions, we can define 
\eeq{
\gamma_{I,ij}(\vec r, z) \equiv  \TF\left[\P_{ik}\P_{jl}~g_{kl}(\vr,z)\right]
= \P_{ijkl} (\vhat r) g_{kl}(\vec r,z) 
\label{eq:gammaI_def}
}
where we defined the total projecting operator 
\eeq{
 \P_{ijkl} (\vhat r) \equiv  \mathcal P_{ik}(\vhat r)\mathcal P_{jl}(\vhat r) - \frac{1}{2} \mathcal P_{ij}(\vhat r)\mathcal P_{kl}(\vhat r) , 
 \label{eq:total_projector}
}
Projector operators have a simple idempotent property $\mathcal P_{ij}\mathcal P_{jk} = \mathcal P_{ik}$ and are  
orthogonal to $ \hat r_i \hat r_j$, but not to the isotropic term $\dK_{ik}$. 

To obtain the Fourier transform of $\gamma_{I,ij}(\vec r, z)$ field the following integral would have to be performed
\eeq{
\gamma_{I,ij}(\vec k, z) = \int d^3r ~e^{i \vec r \cdot \vec k} \gamma_{I,ij}(\vec r, z) = \int \frac{d^3p}{(2\pi)^3}  \mathcal P_{ijkl} (\vec k - \vec p) g_{kl}(\vec p, z).
}
The mode coupling evidenced in the last equality is due to the fact that
the projection operator breaks translation invariance.
This issue is circumvented by performing the spherical harmonic decomposition
on the sky, or the 2D Fourier transform in case when sufficiently small scales are considered (flat-sky approximation).

However, let us first decompose the configuration space tensor field $\gamma_{I,ij}$ into the 
irreducible spherical tensor components, similar to what we did in Fourier space in 
section \ref{sec:galaxy_shapes}. We can thus introduce the helicity basis $(\vhat r,\vec m_+, \vec m_-)$ in analogy to \refeq{k_basis},
\eeq{
\vec m_1 = \frac{\vec e_x \times \v{\hat{r}}}{|\vec e_x \times \v{\hat{r}} |},~~ \vec m_2 = \vhat r \times \vec m_1,~~ 
\vec m^\pm = \mp \tfrac{1}{\sqrt 2} \lb \vec m_1 \mp i \vec m_2 \rb\,,
}
where $\vec e_x$ is a fixed unit vector that is not parallel to $\v{\hat{r}}$,
and construct out of it the rank two harmonic tensor basis 
$\vec M^{(\pm 2)}_{ij} = m_i^{\pm} m_j^{\pm}$. Even though we could introduce the other 
elements of the basis $\vec M^{(0)}_{ij}$ and $\vec M^{(\pm 1)}_{ij}$ (defined in analogy to Eq.~\eqref{eq:Y_basis}) 
these would be orthogonal to the total projection operator $\mathcal P_{ijkl}$ given in Eq.~\eqref{eq:total_projector} 
and thus would not contribute to the decomposition of the projected shape $\gamma_{I,ij}$.
The traceless intrinsic shape field can then be decomposed in terms of two components on the sky
\eeq{
\gamma_{I,ij}( \vec r, z) 
= \gamma_{+2}( \vec r, z)  \vec M^{(+2)}_{ij}(\vhat r) + \gamma_{-2}( \vec r, z)  \vec M^{(-2)}_{ij}(\vhat r).
\label{eq:M_basis}
}
An especially useful property of this basis is the realization that $\vec m^{\pm}$ basis vectors are 
unit eigenfunctions of the projectors $\mathcal P_{ij}$, i.e. we have
\eeq{
\bm{\mathcal P} \cdot \vec m^{\pm} = \vec m^{\pm}.
}
This implies that 
\eeq{
 \vec M^{(\pm2)*}_{ij}(\vhat r) \mathcal P_{ijkl} (\vhat r)
 = m_i^{\mp} m_j^{\mp} \lb \mathcal P_{ik}(\vhat r)\mathcal P_{jl}(\vhat r) - \frac{1}{2} \mathcal P_{ij}(\vhat r)\mathcal P_{kl}(\vhat r) \rb
 =  \vec M^{(\pm2)*}_{kl}(\vhat r),
}
which combined with \refeq{gammaI_def} gives the simple relation  
\eeq{
\gamma_{\pm2}( \vec r, z) = \vec M^{(\pm2)*}_{ij}(\vhat r) g_{ij}(\vec r, z).
\label{eq:gamma_projected_g}
}
The components $\gamma_{\pm2}$ 
transform under rotation as helicity-two functions, 
and we have $\gamma_s^*(\vec r) = (-1)^s \gamma_{-s}(\vec r)$

If we integrate over the line of sight coordinate, 
introducing a window function $W(\chi)$ determined by the redshift distribution $dN(z)/dz$ of galaxies, we have
\eeq{
\label{eq:Wint_all_sky}
\gamma_{\pm2}(\vhat r) \equiv \int d\chi~ W(\chi) \int d^3r' ~ \gamma_{\pm2} \lb \vec r', z[\chi] \rb \dD \lb \vec r' - \chi \vhat r  \rb 
= \int d\chi ~ W(\chi)  \gamma_{\pm2} \big( \chi \vhat r, z[\chi] \big),
}
where the variable $\chi$ is the comoving distance to redshift $z$.

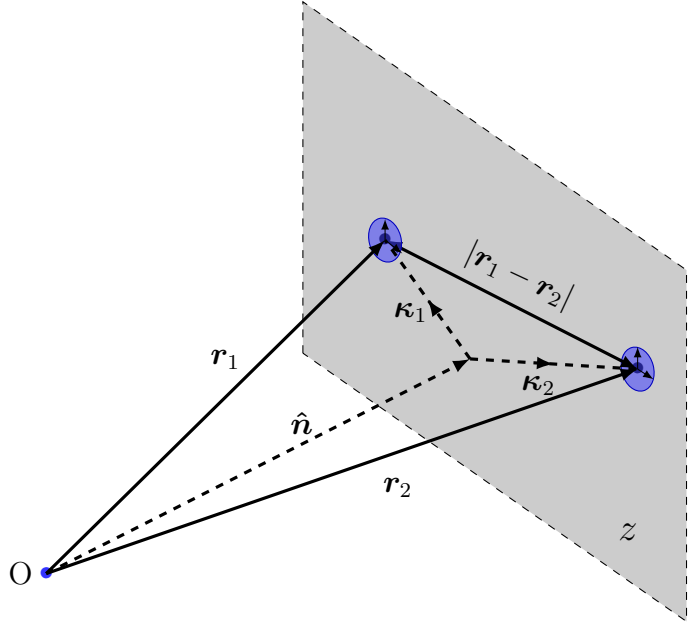
\begin{figure}[t!]
\centering
\begin{tikzpicture}[scale=0.85,every node/.style={minimum size=1cm},on grid]
		    	
    \begin{scope}[
    	xshift=100,every node/.append style={
    	yslant=0.5,xslant=-1},yslant=-0.7,xslant=0.0
    	             ]
    	\fill[gray!80,fill opacity=.5] (-0.5,-0.25) rectangle (5.5,5.25);
    	\draw[black,dashed]  (-0.5,-0.25) rectangle (5.5,5.25);
    \end{scope}
    	    
  \node[blue!80] at (-1,-3.35) {\textbullet}; 
  \node[black!100] at (8.25,-0.15) {\textbullet}; 
  \node[black!100] at (4.3,1.87) {\textbullet}; 
    
  \draw[blue!75!black, fill = blue!75, fill opacity = 0.55, rotate around={15:(8.25,-0.15)}] (8.25,-0.15) ellipse (0.25cm and 0.35cm);
  \draw[->] (8.25,-0.15) -- (8.5,-0.31);
  \draw[->] (8.25,-0.15) -- (8.25,0.2);

  \draw[<->, very thick] (4.3,1.87) -- (8.25,-0.15);

  \draw[blue!75!black, fill = blue!75, fill opacity = 0.55, rotate around={15:(4.3,1.87)}] (4.3,1.87) ellipse (0.25cm and 0.35cm);
  \draw[->] (4.3,1.87) -- (4.57,1.66);
  \draw[->] (4.3,1.87) -- (4.3,2.2);
    
  \draw[->, very thick, dashed] (-1,-3.35) -- (5.63,0.01);
  \draw[->, very thick] (-1,-3.35) -- (8.25,-0.15);
  \draw[->, very thick] (-1,-3.35) -- (4.3,1.87);
  
  \begin{scope}[very thick,decoration={
    markings,
    mark=at position 0.5 with {\arrow{>}}}
    ] 
    \draw[postaction={decorate},dashed]  (5.63,0.01) -- (8.25,-0.15);
    \draw[postaction={decorate},dashed]  (5.63,0.01) -- (4.3,1.87);   
\end{scope}

  \node[black] at (1.8,0.0) {\large $\vec r_1$};
  \node[black] at (4.5,-2.0) {\large $\vec r_2$};
  \node[black] at (3.0,-1.0) {\large $\vhat n$};
  \node[black, rotate=-25] at (6.4,1.3) {\large $| \vec r_{1} - \vec r_{2} |$};
  \node[black] at (4.7,0.74) {\large ${\vec \kappa}_1$};
  \node[black] at (6.7,-0.4) {\large ${\vec \kappa}_2$};
  \node[black] at (-1.4,-3.35) {\large O}; 
  \node[black] at (8.1,-2.7) {\Large $z$};   

\end{tikzpicture}
\caption{Coordinate setup in the flat sky approximation for the angular two-point functions.
We use the coordinates labels where $\vec r =  r_\pp \vhat n + \vec \kappa $, 
with $r_\pp = \vhat n.\vec r$, and $\vec \kappa$ is the coordinate that lies in the observation 
plane, where we introduce the angular variable $\vec \theta = \vec \kappa/\chi$ ($\chi$ is the comoving distance
at redshift $z$). The observer is assumed to be at the point $O$, and the observation plane is assumed 
to be at the distance determined by redshift $z$.}
\label{fig:flat_sky}
\end{figure}

In surveys that do not cover a wide area of the sky, it is often useful and simpler 
to work in the \textit{flat-sky} approximation. Here, the area on the sky is assumed to be 
well approximated by a plane lying perpendicular to a fixed direction $\vhat n$.
This allows us to setup the coordinate frame where $\vec r =  r_\pp \vhat n + \vec \kappa $, 
with $r_\pp = \vhat n.\vec r$, $\vec \kappa = \vec r - r_\pp \vhat n$, and $\vec \theta = \vec \kappa/\chi$. 
This coordinate setup is schematically shown in Figure \ref{fig:flat_sky}. 
In this setting the integral over the redshift is performed along the line of sight $\vhat n$ and we have
\eeq{
\gamma_{\pm 2}(\vec \theta) \equiv \int d\chi~ W(\chi) \int d^3r' ~  \gamma_{\pm 2} \big( \vec r', z[\chi] \big) \dD \lb \vec r ' - \chi \vhat n - \vec \kappa  \rb 
= \int d\chi ~ W(\chi)  \gamma_{\pm2} \big( \chi \vhat n , \chi \vec \theta, z[\chi] \big),
}
and the spherical tensors $ \vec M^{(\pm2)}_{ij}(\vhat n)$ are also defined relative to the $\vhat n$ direction. 
Since we have established a 2D planar coordinate $\vec \theta$, we can also introduce a corresponding 2D 
Fourier transform. This is again useful, in the flat sky approximation, since the statistical isotropy 
can serve as an homogeneity condition on the plane and we can thus use statistical translation invariance. 
We thus define a Fourier field on a plane
\eq{
\gamma_{I,ij}(\vec \ell) = \int d^2\vec \theta~ \gamma_{I,ij}(\vec \theta) e^{i \vec \ell . \vec \theta}.
}
and the decomposition in spherical tensors defined in the $\pmb \ell$ plane orthogonal to the line of sight $\v{\hat{r}}$ is
\eeq{
\gamma_{I,ij}(\pmb \ell) = \gamma_{+2}(\pmb \ell)  \vec{ \widetilde M}^{(+2)}_{ij} (\vhat n) + \gamma_{-2}(\pmb \ell)  \vec{ \widetilde M}^{(-2)}_{ij}(\vhat n).
}
The basis $\vec{ \widetilde M}^{(\pm2)}$ spans a 2D Fourier space while orthogonal to the $\vhat n$ direction. 
We can think of it as being constructed from vector basis  $(\vhat n,\vhat \ell_+, \vhat \ell_-)$, analogously as we did in \refeq{k_basis}, but with $\vhat \ell_\pm$ referring to a fixed direction $\vhat n$.
Each of the helicity components in this Fourier space is given by
\eeq{
 \gamma_{\pm2}(\vec \ell)  = \vec{ \widetilde M}^{(\pm2)*}_{ij} (\vhat n)  \gamma_{I,ij}(\vec \ell) 
 = \int d^2\vec \theta ~ \gamma_{\pm2}(\vec \theta) e^{\pm2i(\phi_\theta - \phi_\ell)} e^{i \vec \ell . \vec \theta}.
 \label{eq:gamma_2D_FT}
}
Note that it is not obvious that $\gamma_{\pm2}(\vec \ell) $, as defined above, will correspond
to the small-angle limit of all sky treatment for high enough $\ell$. However, this correspondence has been 
explicitly established in e.g. \cite{Hu:2000}, by taking the small-angle limit of the all sky results, 
and we will return to it in our upcoming \textit{all-sky} paper. 

Very frequently, different bases for decomposing the $\gamma_{I,ij}$ are used in the literature. Some of the most prominent are
the \textit{polarization basis} $(E,B)$ used in the study of the CMB polarization \cite{Seljak:1996, Zaldarriaga+:1996, Hu+:1997},
and what we label the \textit{Pauli basis} $(\gamma_1, \gamma_2)$. 
The relation of our helicity basis to the polarization basis of galaxy shapes $(\gamma_E,\gamma_B)$ is a simple linear combination 
\eeq{
\gamma_E = \frac{1}{2} \lb \gamma_{+2} + \gamma_{-2} \rb , ~~ \gamma_B = -\frac{i}{2} \lb \gamma_{+2} - \gamma_{-2} \rb,
\label{eq:polarization_gamma}
}
with the inverse $\gamma_{\pm2} = \lb \gamma_E \pm i \gamma_B \rb$. Since we deal with the simple linear combination 
of the fields, these relations hold in both all-sky and flat-sky, as well as in both angle and Fourier space of 
the flat-sky approximation. 
On the other hand Pauli basis can be obtained by similar linear combination and additional rotation 
around the azimuthal angle $\phi$. We have
\eeq{
\gamma_1 =  \frac{1}{2} \lb \gamma_{+2} e^{2i \phi} + \gamma_{-2} e^{-2i \phi}  \rb,~~
\gamma_2 =  \frac{i}{2} \lb \gamma_{+2} e^{2i \phi} - \gamma_{-2} e^{-2i \phi}  \rb.
}
and the inverse is given by $\gamma_{\pm2} = e^{\pm 2i \phi}  \lb \gamma_1 \pm i \gamma_2 \rb$, where $\phi = \phi_\theta$ in real space and $\phi = \phi_\ell$ in Fourier space, denoting the azimuthal angles of $\v{\theta}$ and $\v{\ell}$, respectively. 
The Pauli basis can again be used either in all-sky or flat-sky, however it is more commonly used
in the flat-sky case. As the name suggests, on the 2D sky in this basis can be organized 
as the linear combination 
\eeq{
  g_{ij}(\vec\theta) =
\begin{pmatrix} 
\gamma_1 & \gamma_2 \\
\gamma_2 & - \gamma_1
\end{pmatrix} _{ij}
= \gamma_1(\vec\theta)  (\sigma_3)_{ij} +  \gamma_2(\vec\theta) (\sigma_1)_{ij},
}
where  $\vec \sigma_1$ and $\vec \sigma_3$ are the Pauli matrices (see e.g. \cite{Castro++:2005}).
The relation of the Pauli basis and polarization basis is related by the simple plane rotation by a $2\phi_\theta$ spin angle. 
Electric ($E$) and magnetic ($B$) components in angle space are then given by
\eq{
\label{eq:Pauli_polarization_basis}
\gamma_E(\vec \theta)& = \cos(2 \phi_\theta) \gamma_1(\vec \theta)  + \sin(2 \phi_\theta)  \gamma_2(\vec \theta), \\
\gamma_B(\vec \theta)& = -\sin(2 \phi_\theta) \gamma_1(\vec \theta)  + \cos(2 \phi_\theta)  \gamma_2(\vec \theta), \non
}
and equivalent expressions, as already noted, hold in the 2D Fourier space.

\section{Angular power spectra}
\label{sec:flat-sky}

In this part, we explore the two- and three-point statistics in 
the flat-sky approximation. The coordinate set-up is as described 
in the earlier section and as suggested for the two-point function in
Figure \ref{fig:flat_sky}. These results agree with the corresponding 
all-sky result in the small-angle (high $\ell$) limit, as will be 
shown in an upcoming paper. 

\subsection{Two-point functions}

We can evaluate the angular correlation function using \refeq{gamma_2D_FT}.
We are interested in the following correlators:
\eq{
\la \df_{\rm n}(\vec \ell_1) | \df_{\rm n}^*(\vec \ell_2) \ra 
&= \int d^2 \vec \theta_1 d^2 \vec \theta_2 ~ \la \df_{\rm n}(\vec \theta_1) \df_{\rm n}^*(\vec \theta_2) \ra e^{i \vec \ell_1 . \vec \theta_1 - i \vec \ell_2 . \vec \theta_2}, \\
\la \df_{\rm n}(\vec \ell_1) | \gamma^*_{s_2} (\vec \ell_2) \ra 
&= \int d^2 \vec \theta_1 d^2 \vec \theta_2 ~ \la \df_{\rm n}(\vec \theta_1) \gamma^*_{s_2}(\vec \theta_2) \ra e^{ - i s_2 (\phi_{\theta_2} - \phi_{\ell_2})} 
e^{i \vec \ell_1 . \vec \theta_1 - i \vec \ell_2 . \vec \theta_2}, \non\\
\la \gamma_{s_1} (\vec \ell_1) | \gamma^*_{s_2} (\vec \ell_2) \ra 
&= \int d^2 \vec \theta_1 d^2 \vec \theta_2 ~ \la \hat  \gamma_{s_1}(\vec \theta_1) \gamma^*_{s_2}(\vec \theta_2) \ra 
e^{ i s_1 (\phi_{\theta_1} - \phi_{\ell_1}) - i s_2 (\phi_{\theta_2} - \phi_{\ell_2})} e^{i \vec \ell_1 . \vec \theta_1 - i \vec \ell_2 . \vec \theta_2}. \non
}
We use the $s$ variable to label the helicity states $\pm 2$, and we will refer to $\tilde s$ as the sign of the helicity state, i.e. $\tilde s=s/|s|$.
Here and throughout, we assume galaxy number counts $\df_{\rm n}$ as the scalar observable we are correlating with. The expressions however are applicable generally to any such scalar (e.g., galaxy sizes $\df_{\rm s}$).

We note again that in the flat-sky approximation statistical isotropy manifests itself as the 
translation invariance (and rotation invariance around the line of sight) of the correlators 
on the plane. This allows us to define simple spectral functions that depend only on the 
amplitude of the $\vec \ell$-mode on the plane
\eq{
\label{eq:flat_sky_angluar_ps}
 (2\pi)^2 \dD(\vec \ell - \vec \ell')C_{{\rm n} {\rm n}}(\ell) &= \la \df_{\rm n}(\vec \ell) | \df_{\rm n}^*(\vec \ell') \ra, \\ 
 (2\pi)^2 \dD(\vec \ell - \vec \ell')C_{{\rm n} \pm}(\ell) &= \la \df_{\rm n}(\vec \ell) | \gamma^*_{\pm2}(\vec \ell') \ra, \non\\ 
 (2\pi)^2 \dD(\vec \ell - \vec \ell')C_{\pm\pm}(\ell) &= \la \gamma_{\pm2}(\vec \ell) | \gamma^*_{\pm2}(\vec \ell') \ra. \non
}
The explicit calculation of these angular power spectra is given in Appendix 
\ref{app:flat_sky_spectrum}. 
We use the standard approximation that the longitudinal modes $k_\pp = \vhat n . \vk$ do not 
contribute noticeably to the angular spectra compared to the modes perpendicular to the plane, i.e. $k_\pp \ll k_\perp$.
Note that this approximation is not strictly necessary, it however becomes
accurate in the same regime where the flat-sky approximation does,
and greatly simplifies the resulting expressions.
Using the resulting expressions given in \refeq{flat_sky_angular_Cl} we obtain
\eq{
\label{eq:flat_sky_angular_Cl_main}
C_{{\rm n} +}(\ell) = C_{{\rm n} -}(\ell) &= \frac{1}{2}  \int d\chi  ~ \frac{W_{\rm n}(\chi)W_g(\chi)}{\chi^2}~ N_0 P_{02}^{(0)}(\ell/ \chi), \\
C_{++}(\ell) = C_{--}(\ell) &= \frac{1}{8} \int d\chi  ~ \frac{W_g^2(\chi)}{\chi^2}  \lb 2 N_0^2  P_{22}^{(0)}( \ell/ \chi) + 8 N_1^2  P_{22}^{(1)}( \ell/ \chi) + N_2^2  P_{22}^{(2)}( \ell/ \chi) \rb, \non\\
C_{+-}(\ell) = C_{-+}(\ell) &= \frac{1}{8} \int d\chi  ~ \frac{W_g^2(\chi)}{\chi^2}  \lb 2 N_0^2  P_{22}^{(0)}( \ell/ \chi) - 8 N_1^2  P_{22}^{(1)}( \ell/ \chi) + N_2^2  P_{22}^{(2)}( \ell/ \chi) \rb, \non
}
where $W_{\rm n}$ and $W_{\rm g}$ denote the window functions corresponding to number counts and shapes, respectively, and 
we have the normalization constants $N_{0,1,2}=\left\{ \sqrt{\tfrac{3}{2}}, \sqrt{\frac{1}{2}}, 1\right\}$ 
(related to the basis $\vec Y^{(m)}_{ij}$ in Eq.~\ref{eq:Y_basis}). 
Above, we allowed for the fact that the window functions for number density $W_{\rm n}$ and for shapes $W_g$ can be different. 
Note that when using these flat-sky expressions,
it is found to be a much better approximation to use $\ell \to \ell +1/2$ in the arguments of $P^{q}_{s s'}(\ell/\chi)$ 
power spectra above (see e.g. \cite{LoVerde+:2008}). With this correction, the estimated error of the approximation
is reduced from $O(\ell^{-1})$ to $O(\ell^{-2})$.

We can also consider the angular spectra in different bases, introduced earlier. 
In the polarization basis, defined in \refeq{polarization_gamma}, angular power spectra are
\eq{
C_{{\rm n} E}(\ell) &= \frac{1}{2}  \int d\chi  \frac{W_{\rm n}(\chi)W_g(\chi)}{\chi^2} N_0 P^{(0)}_{02} ( \ell / \chi), \label{eq:Cl_flatsky}\\
C_{EE}(\ell) &= \frac{1}{8} \int d\chi  \frac{W_g^2(\chi)}{\chi^2} \lb 2 N_0^2 P^{(0)}_{22} ( \ell / \chi) + N_2^2 P^{(2)}_{22} ( \ell / \chi) \rb \non\\
C_{BB}(\ell) &= \int d\chi  \frac{W_g^2(\chi)}{\chi^2} N_1^2 P^{(1)}_{22} ( \ell / \chi). \non
}
The cross correlators of $B$-modes with $E$-modes 
or a scalar field vanish as a consequence of statistical parity invariance (see e.g. Paper I \cite{Vlah++:2019}): $C_{{\rm n} B}(\ell)= C_{EB}(\ell) = 0$. 
The angular power spectra in the helicity basis $\gamma_\pm$ and the polarization basis 
are related by
\eq{
C_{{\rm n} +}(\ell) = C_{{\rm n} -}(\ell) &= C_{{\rm n} E}(\ell), \\
C_{++}(\ell) = C_{--}(\ell) &= C_{EE}(\ell) + C_{BB}(\ell) , \non\\
C_{+-}(\ell) = C_{-+}(\ell) &= C_{EE}(\ell) - C_{BB}(\ell). \non
}
The Pauli basis is related to the polarization by simple rotation in the plane \refeq{Pauli_polarization_basis}.
Angular power spectra in this basis are
\eq{
C_{{\rm n}\gamma_1}(\ell,\phi_\ell) &= \cos(2 \phi_\ell) C_{{\rm n} E}(\ell), \\
C_{{\rm n}\gamma_2}(\ell,\phi_\ell) &= \sin(2 \phi_\ell) C_{{\rm n} E}(\ell), \non\\
C_{\gamma_1\gamma_1}(\ell,\phi_\ell) &= \cos(2 \phi_\ell)^2 C_{EE}(\ell) + \sin(2 \phi_\ell)^2 C_{BB}(\ell), \non\\
C_{\gamma_1\gamma_2}(\ell,\phi_\ell) &= \sin(2 \phi_\ell)\cos(2 \phi_\ell)  \lb C_{EE}(\ell) - C_{BB}(\ell) \rb, \non\\
C_{\gamma_2\gamma_2}(\ell,\phi_\ell) &= \sin(2 \phi_\ell)^2 C_{EE}(\ell) + \cos(2 \phi_\ell)^2 C_{BB}(\ell). \non
}

So far, all expressions included intrinsic contributions to shape correlations.
The corresponding expressions including the leading lensing effect are given
in Appendix \ref{app:lensing}.

Let us now return to the issue of nonlinear projection contributions.
Since the shape measurement process is nonlinear, and the galaxy shape
field is weighted by the number density of source galaxies, \refeq{g_projected}
receives additional corrections. One such effect is that shapes are in fact
measuring the reduced shear \cite{Bartelmann+:1999,White:2005,Dodelson++:2005}.
However there are also nonlinear effects which depend on the intrinsic shapes
of galaxies \cite{Schmidt++_I:2009,Schmidt++_II:2009,Krause++:2009}.
Some examples of resulting contributions are
\be
\gamma_{ij}(\vr^s,z) \supset \delta_g \gamma_{G,ij}, \quad
\kappa \TF\left[\P_{ik}\P_{jl} ~ g_{kl}(\vr[\vr^s],z)\right], \quad
\gamma_{G,(im} \TF\left[\P_{mk}\P_{j)l} ~ g_{kl}(\vr[\vr^s],z)\right].
\label{eq:g_projected_nl}
\ee
where $\gamma_G$ denotes the lensing contribution to the shape distortion. 
On the other hand, in terms of purely intrinsic contributions, the effect of
number-density weighting is already included in our shape bias expansion,
as explained in Paper I \cite{Vlah++:2019}. The contributions in \refeq{g_projected_nl} schematically add terms of the following form to angular shear correlations: 
\be
C_{EE}(\ell) \Big|_\text{nonl. proj.} \propto \int \frac{d^2 \ell'}{(2\pi)^2}
C^I_{EE}(\ell') C^G_{EE}(|\ell-\ell').
\ee
That is, the convolution is performed after the projection. 
This is in contrast to the 1-loop shape contributions in \refeq{Cl_flatsky},
which involve the projection of a 3D convolution integral. 
As shown in detail in \cite{Jeong++:2011}, the latter contributions
are much larger than the contributions from nonlinear projection effects. 
For this reason, we are justified in ignoring the latter here, even though
they are formally of the same order in perturbations.

In terms of 
lensing contributions only, this hierarchy is well known: reduced-shear
corrections to the shear power spectrum are much smaller than the
corrections to the nonlinearities in the matter density field
\cite{Dodelson++:2005,Schmidt++_II:2009,Krause++:2009}.

\subsection{Three-point functions}

We turn next to the angular bispectrum calculation in the flat-sky approximation.
The details of the derivation are presented in Appendix \ref{app:flat_sky_bispectrum}
and here we give a summary of the results. 

In analogy to how we define the flat-sky angular power spectrum in \refeq{flat_sky_angluar_sin} we can define the angular bispectrum.
We thus have
\eq{
\label{eq:flat_sky_angluar_sin}
 (2\pi)^2 \dD \lb \vec \ell_1 + \vec \ell_2 +\vec \ell_3  \rb B_{{\rm n} {\rm n} {\rm n}} \lb \vec \ell_1, \vec \ell_2, \vec \ell_3 \rb &= \la \df_{\rm n}(\vec \ell_1) \df_{\rm n}(\vec \ell_2) \df_{\rm n}(\vec \ell_3) \ra, \\ 
 (2\pi)^2 \dD \lb \vec \ell_1 + \vec \ell_2 +\vec \ell_3  \rb B_{{\rm n} {\rm n} \pm} \lb \vec \ell_1, \vec \ell_2, \vec \ell_3 \rb &= \la \df_{\rm n}(\vec \ell_1) \df_{\rm n}(\vec \ell_2) \gamma_{\pm2}(\vec \ell_3) \ra, \non\\ 
 (2\pi)^2 \dD \lb \vec \ell_1 + \vec \ell_2 +\vec \ell_3  \rb B_{{\rm n} \pm \pm} \lb \vec \ell_1, \vec \ell_2, \vec \ell_3 \rb &= \la \df_{\rm n}(\vec \ell_1) \gamma_{\pm2}(\vec \ell_2) \gamma_{\pm2}(\vec \ell_3)\ra, \non\\ 
 (2\pi)^2 \dD \lb \vec \ell_1 + \vec \ell_2 +\vec \ell_3  \rb B_{\pm \pm \pm} \lb \vec \ell_1, \vec \ell_2, \vec \ell_3 \rb &= \la \gamma_{\pm2}(\vec \ell_1)\gamma_{\pm2}(\vec \ell_2) \gamma_{\pm2}(\vec \ell_3)\ra. \non
}
To obtain the relation of these angular bispectra to the 3D spherical tensor bispectra in Section \ref{sec:three-point functions_in_3D} in the flat-sky approximation
we again use the approximation that the longitudinal modes $k_\pp = \vhat n . \vk$ contribute negligibly to the perpendicular modes $k_\perp$ that lay in the plane 
of the sky (see Figure \ref{fig:flat_sky}).  We thus assume that $k_\pp \ll k_\perp$. We again note that relaxing this approximation would give rise to the subleading corrections in $\vec \ell$ variables.
However, with this approximation we ensure translation invariance in the plane manifested as the Dirac delta's 
on the left hand side of definitions in \refeq{flat_sky_angluar_sin}.
We obtain the results given in \refeq{angular_bispectrum_flat_sky_app} that we review here. 
The density-density-density angular bispectrum is given by
\eq{
B_{{\rm n}{\rm n}{\rm n}} \lb \vec \ell_1, \vec \ell_2, \vec \ell_3 \rb & = \int d\chi ~ \frac{[W_{\rm n}(\chi)]^3}{\chi^4}  B^{(0)}_{000} \lb \tilde{\vec \ell}_1, \tilde{\vec \ell}_2, \tilde{\vec \ell}_3 \rb, 
}
where we use the shorthand notation $\tilde{\vec \ell} = \vec \ell \chi$.  This result is the standard clustering bispectrum \cite{Bartelmann+:1999}.

The density-density-shape angular bispectrum is given by
\eq{
B_{{\rm n}{\rm n} \pm} \lb \vec \ell_1, \vec \ell_2, \vec \ell_3 \rb & = \frac{1}{2} \int d\chi ~ \frac{[W_{\rm n}(\chi)]^2 W_g(\chi)}{\chi^4}
\bigg[ N_0 B^{(0)}_{002} \mp i 2 \sqrt{2} N_1 B^{(1)}_{002}  - N_2 B^{(2)}_{002} \bigg] \lb \tilde{\vec \ell}_1, \tilde{\vec \ell}_2, \tilde{\vec \ell}_3 \rb.
}
In the polarization basis introduced in \refeq{polarization_gamma}, density-density-shape angular bispectrum can be 
split into $E$ and B components yielding
\eq{
\label{eq:bis_nnE_nnB}
B_{{\rm n}{\rm n} E} \lb \vec \ell_1, \vec \ell_2, \vec \ell_3 \rb & = 
\frac{1}{2} \int d\chi ~ \frac{[W_{\rm n}(\chi)]^2 W_g(\chi)}{\chi^4}
\bigg[ N_0 B^{(0)}_{002}- N_2 B^{(2)}_{002} \bigg] \lb \tilde{\vec \ell}_1, \tilde{\vec \ell}_2, \tilde{\vec \ell}_3 \rb, \\
B_{{\rm n}{\rm n} B} \lb \vec \ell_1, \vec \ell_2, \vec \ell_3 \rb & =
- \sqrt{2} N_1 \int d\chi ~ \frac{[W_{\rm n}(\chi)]^2 W_g(\chi)}{\chi^4}  B^{(1)}_{002}  \lb \tilde{\vec \ell}_1, \tilde{\vec \ell}_2, \tilde{\vec \ell}_3 \rb. \non
}
This decomposition has already been shown in Paper I \cite{Vlah++:2019}
(see also \cite{Schmitz++:2018} for similar results). 

The density-shape-shape angular bispectrum in helicity basis can compactly be written as
\eq{
B_{{\rm n} \tilde s_2 \tilde s_3} \lb \vec \ell_1, \vec \ell_2, \vec \ell_3 \rb & =  \frac{1}{8} \int d\chi ~ \frac{W_{\rm n}(\chi) [W_g(\chi)]^2}{\chi^4}
\bigg[
2 N_0^2  B^{(0,0)}_{022} 
+ 2 s_2 s_3  N_1^2  \lb B^{(-1,1)}_{022} - B^{(1,1)}_{022} \rb \\
&\hspace{3.cm} - i 2 \sqrt{2} N_0 N_1 \lb s_3 B^{(0,1)}_{022} + s_2 B^{(1,0)}_{022} \rb \non\\
&\hspace{3.cm} + i \sqrt{2} N_1 N_2 \lb s_2 B^{(1,2)}_{022} + s_3 B^{(2,1)}_{022} + s_3 B^{(-2,1)}_{022} + s_2 B^{(1,-2)}_{022} \rb \non\\
&\hspace{3.cm} - 2 N_0 N_2  B^{(\{0,2\})}_{022} 
+ N_2^2  \lb B^{(-2,2)}_{022} + B^{(2,2)}_{022} \rb 
\bigg]  \lb \tilde{\vec \ell}_1, \tilde{\vec \ell}_2, \tilde{\vec \ell}_3 \rb . \non
}
It is important to notice that 
if we consider the cross-bispectrum (correlating at least two different shape tracers)
we have four independent angular bispectra, since $B_{{\rm n} - +}$ can be different from $B_{{\rm n} + -}$.
In the case when only auto-bispectra (same tracers) are considered this simplifies and we have 
$B_{{\rm n} - +}=B_{{\rm n} + -}$. This can be seen if we consider the difference of the two angular 
bispectra. Using the explicit form above we have 
\eq{
B_{{\rm n} - +} - B_{{\rm n} + -} &= - \frac{i}{\sqrt 2} N_1 \int d\chi ~ \frac{W_{\rm n}(\chi) [W_g(\chi)]^2}{\chi^4}
\bigg[
2 N_0 \lb B^{(1,0)}_{022} - B^{(0,1)}_{022}\rb \\
&\hspace{5.0cm} + N_2 \lb B^{(1,-2)}_{022} - B^{(-2,1)}_{022} + B^{(1,2)}_{022} - B^{(2,1)}_{022}\rb
\bigg], \non
}
which vanishes if the auto-correlation bispectrum is considered. 
This situation is intrinsically different from the case of power spectrum where $C_{+-} = C_{-+}$ 
(see Eq.~\ref{eq:flat_sky_angular_Cl_main}) was guaranteed even for the cross correlations 
of different shape tracers. The reason for this lies in the fact that the imposed symmetries (statistical 
isotropy and parity invariance) are more constraining for the two-point functions, leaving less functional 
freedom than in case of three-point functions.

We can transform these results into the $E$ and $B$ basis to get 
\eq{
\label{eq:bis_022_EB}
B_{{\rm n} E E}\lb \vec \ell_1, \vec \ell_2, \vec \ell_3 \rb &= \int d\chi ~ \frac{W_{\rm n}(\chi) [W_g(\chi)]^2}{4 \chi^4}
\bigg[ N_0^2 B^{(0,0)}_{022} + N_0 N_2 B^{(\{0,2\})}_{022} \\
&\hspace{5.8cm} + \frac{N_2^2}{2} \lb B^{(-2,2)}_{022} + B^{(2,2)}_{022} \rb \bigg]  \lb \tilde{\vec \ell}_1, \tilde{\vec \ell}_2, \tilde{\vec \ell}_3 \rb, \non\\
B_{{\rm n} EB}\lb \vec \ell_1, \vec \ell_2, \vec \ell_3 \rb &= N_1 \int d\chi ~ \frac{W_{\rm n}(\chi) [W_g(\chi)]^2}{2\sqrt 2 \chi^4} 
\bigg[ - 2N_0 B^{(0,1)}_{022} \non\\
&\hspace{5.8cm} + N_2 \lb B^{(-2,1)}_{022} + B^{(2,1)}_{022} \rb  \bigg]  \lb \tilde{\vec \ell}_1, \tilde{\vec \ell}_2, \tilde{\vec \ell}_3 \rb, \non\\
B_{{\rm n} BE}\lb \vec \ell_1, \vec \ell_2, \vec \ell_3 \rb &= N_1 \int d\chi ~ \frac{W_{\rm n}(\chi) [W_g(\chi)]^2}{2\sqrt 2 \chi^4}
\bigg[ - 2N_0 B^{(1,0)}_{022} \non\\
&\hspace{5.8cm} + N_2 \lb B^{(1,-2)}_{022} + B^{(1,2)}_{022} \rb \bigg]  \lb \tilde{\vec \ell}_1, \tilde{\vec \ell}_2, \tilde{\vec \ell}_3 \rb, \non\\
B_{{\rm n} BB}\lb \vec \ell_1, \vec \ell_2, \vec \ell_3 \rb &= N_1^2 \int d\chi ~ \frac{W_{\rm n}(\chi) [W_g(\chi)]^2}{\chi^4} 
\bigg[ B^{(1,1)}_{022} + B^{(-1,1)}_{022} \bigg]  \lb \tilde{\vec \ell}_1, \tilde{\vec \ell}_2, \tilde{\vec \ell}_3 \rb. \non 
}
We again note that $B_{{\rm n} EB}$ and $B_{{\rm n} BE}$ are equal in the case of tracer auto-correlations, while in general they can be different when cross-correlating different tracers.

Finally, we turn to the correlations of the three shape fields. The general expression in terms of the helicity 
basis is given by 
\eq{
B_{\tilde s_1\tilde s_2 \tilde s_3} \lb \vec \ell_1, \vec \ell_2, \vec \ell_3 \rb & =  \frac{1}{32}  \int d\chi ~ \frac{[W_g(\chi)]^3}{\chi^4}
\bigg[
4 N_0^3  B^{(0,0,0)}_{222}  - 4 N_0^2 N_2 B^{(\{0,0,2\})}_{222} \\
&\hspace{0.5cm} - N_2^3 \lb B^{(\{-2,2,2\})}_{222} + B^{(2,2,2)}_{222} \rb \non\\
&\hspace{0.5cm} + 4 N_0 N_1^2 \lb s_2 s_3 \lb B^{(0,-1,1)}_{222} - B^{(0,1,1)}_{222} \rb + 2~{\rm cyc. }  \rb \non\\
&\hspace{0.5cm} + 2 N_0 N_2^2 \lb B^{(\{0,-2,2\})}_{222} + B^{(\{0,2,2\})}_{222} \rb \non\\
&\hspace{0.5cm} + 2 N_1^2 N_2 \lb s_1 s_2 \lb B^{(1,1,2)}_{222} + B^{(1,1,-2)}_{222} - B^{(\{-1,1\},2)}_{222} \rb  + 2~{\rm cyc. } \rb \non\\
&\hspace{0.5cm} - i 4 \sqrt{2} N_0^2 N_1 \lb s_3 B^{(0,0,1)}_{222} + 2~{\rm cyc. }  \rb \non\\
&\hspace{0.5cm} - i 2 \sqrt{2} s_1s_2s_3 N_1^3 \lb B^{(\{-1,1,1\})}_{222} - B^{(1,1,1)}_{222} \rb \non\\
&\hspace{0.5cm} - i \sqrt{2} N_1 N_2^2 \lb s_3 \lb B^{(\{-2,2\},1)}_{222} + B^{(2,2,1)}_{222} - B^{(2,2,-1)}_{222} \rb + 2~{\rm cyc. }  \rb \non\\
&\hspace{0.5cm} + i 2 \sqrt{2} N_1 N_2 N_3 \lb s_3 \lb B^{(\{0,-2\},1)}_{222} + B^{(\{0,2\},1)}_{222} \rb + 2~{\rm cyc. }  \rb \bigg]   \lb \tilde{\vec \ell}_1, \tilde{\vec \ell}_2, \tilde{\vec \ell}_3 \rb. \non
}
As in case of the density shape-shape bispectra, we see that in the case where we are considering the single tracer auto-correlations, 
we have $B_{++-}=B_{+-+}=B_{-++}$, and similarly for the permutations of $B_{+--}$.
Transforming these results into the $E$ and $B$ basis, we get 
\eq{
\label{eq:bis_222_EB}
B_{EEE}\lb \vec \ell_1, \vec \ell_2, \vec \ell_3 \rb &= \int d\chi ~ \frac{W_{\rm n}(\chi) [W_g(\chi)]^2}{32 \chi^4}
\bigg[ 4N_0^3 B^{(0,0,0)}_{222} - 4N_0^2 N_2 B^{(\{0,0,2\})}_{222} \\
&\hspace{5.7cm} + 2 N_0 N_2^2 \lb B^{(\{0,-2,2\})}_{222} - B^{(\{0,2,2\})}_{222} \rb \non\\
&\hspace{5.7cm} - N_2^3 \lb B^{(\{-2,2,2\})}_{222} + B^{(2,2,2)}_{222} \rb
 \bigg]  \lb \tilde{\vec \ell}_1, \tilde{\vec \ell}_2, \tilde{\vec \ell}_3 \rb, \non\\
B_{EEB}\lb \vec \ell_1, \vec \ell_2, \vec \ell_3 \rb &= - N_1 \int d\chi ~ \frac{W_{\rm n}(\chi) [W_g(\chi)]^2}{8\sqrt 2 \chi^4} 
\bigg[ 4 N_0^2 B^{(0,0,1)}_{222} - 2 N_0 N_2 \lb B^{(\{0,-2\},1)}_{222} + B^{(\{0,2\},1)}_{222} \rb \non\\
&\hspace{3.8cm} + N_2^2 \lb B^{(\{-2,2\},1)}_{222} + B^{(2,2,1)}_{222} - B^{(2,2,-1)}_{222} \rb  \bigg]  \lb \tilde{\vec \ell}_1, \tilde{\vec \ell}_2, \tilde{\vec \ell}_3 \rb, \non\\
B_{EBB}\lb \vec \ell_1, \vec \ell_2, \vec \ell_3 \rb &= N_1^2 \int d\chi ~ \frac{W_{\rm n}(\chi) [W_g(\chi)]^2}{4 \chi^4}
\bigg[ 2N_0 \lb  B^{(0,1,1)}_{222} - B^{(0,-1,1)}_{222} \rb \non\\
&\hspace{3.8cm} + N_2 \lb B^{(2,\{-1,1\})}_{222} - B^{(2,1,1)}_{222} - B^{(-2,1,1)}_{222} \rb \bigg]  \lb \tilde{\vec \ell}_1, \tilde{\vec \ell}_2, \tilde{\vec \ell}_3 \rb, \non\\
B_{BBB}\lb \vec \ell_1, \vec \ell_2, \vec \ell_3 \rb &= N_1^3 \int d\chi ~ \frac{W_{\rm n}(\chi) [W_g(\chi)]^2}{\sqrt{2}\chi^4} 
\bigg[ B^{(\{-1,1,1\})}_{222} - B^{(1,1,1)}_{222} \bigg]  \lb \tilde{\vec \ell}_1, \tilde{\vec \ell}_2, \tilde{\vec \ell}_3 \rb. \non 
}
Where, to obtain the values, for other cross-corrlation components, like $B_{EBE}$ etc., 
we need to take permutations of the $q=1$ helicity index in $B^{(q_1,q_2,q_3)}_{222}$ bispectra in the expressions above.
In case of auto correlations these are all equivalent again.

In Paper I \cite{Vlah++:2019}, we presented the results for the 3D bispectra  $B^{(q_1,q_2,q_3)}_{\ell_1,\ell_2,\ell_3}$
at tree level (LO - leading order) PT. We also showed explicitly the results for the density-density-shape angular 
bispectra, demonstrating that $B_{{\rm n}{\rm n} E}$ both $B_{{\rm n}{\rm n} B}$ contributions arise already at leading PT order (see also \cite{Schmitz++:2018}). 
For density-shape-shape and shape-shape-shape angular bispectra, the situation is different. At tree level, only one nonzero helicity $q$ can contribute. 
This is because at leading order in PT, the $\Pi_{ij}(\vr)$ field given in \refeq{hatPi} has only scalar contributions. 
The contributions that survive at leading order for all bispectra are then listed in Eq.\eqref{eq:LO_bis} of  Appendix \ref{app:flat_sky_bispectrum}. 
We see that the $B_{{\rm n}{\rm n} E}$ and $B_{{\rm n}{\rm n} B}$ expressions given in Eq.~\eqref{eq:bis_nnE_nnB}
remain unchanged, while for the rest the number of contributions is substantially reduced.
The expressions given in \refeq{bis_022_EB} at tree level thus reduce to
\eq{
B_{{\rm n} EE}\lb \vec \ell_1, \vec \ell_2, \vec \ell_3 \rb &= N_0 \int d\chi ~ \frac{W_{\rm n}(\chi) [W_g(\chi)]^2}{4 \chi^4} \bigg[ N_0 B^{(0,0)}_{022} + N_2 B^{(\{0,2\})}_{022} \bigg]  \lb \tilde{\vec \ell}_1, \tilde{\vec \ell}_2, \tilde{\vec \ell}_3 \rb, \\
B_{{\rm n} EB}\lb \vec \ell_1, \vec \ell_2, \vec \ell_3 \rb &= - N_0 N_1 \int d\chi ~ \frac{W_{\rm n}(\chi) [W_g(\chi)]^2}{\sqrt 2 \chi^4} B^{(0,1)}_{022} \lb \tilde{\vec \ell}_1, \tilde{\vec \ell}_2, \tilde{\vec \ell}_3 \rb, \non\\
B_{{\rm n} BB}\lb \vec \ell_1, \vec \ell_2, \vec \ell_3 \rb &= 0, \non 
}
where $B_{{\rm n} BE}$ is again equal to $B_{{\rm n} EB}$ in case of auto-correlations, while in case of cross-correlations 
in the expression above we need to replace the $B^{(0,1)}_{022}$ with $B^{(1,0)}_{022}$.
Similarly, from \refeq{bis_222_EB} we get a substantial reduction of contributions
\eq{
B_{EEE}\lb \vec \ell_1, \vec \ell_2, \vec \ell_3 \rb &=  N_0^2 \int d\chi ~ \frac{W_{\rm n}(\chi) [W_g(\chi)]^2}{8 \chi^4}
\bigg[ N_0 B^{(0,0,0)}_{222} - N_2 B^{(\{0,0,2\})}_{222} \bigg]  \lb \tilde{\vec \ell}_1, \tilde{\vec \ell}_2, \tilde{\vec \ell}_3 \rb, \\
B_{EEB}\lb \vec \ell_1, \vec \ell_2, \vec \ell_3 \rb &= - N_0^2 N_1 \int d\chi ~ \frac{W_{\rm n}(\chi) [W_g(\chi)]^2}{2\sqrt 2 \chi^4} B^{(0,0,1)}_{222} \lb \tilde{\vec \ell}_1, \tilde{\vec \ell}_2, \tilde{\vec \ell}_3 \rb, \non\\
B_{EBB}\lb \vec \ell_1, \vec \ell_2, \vec \ell_3 \rb &= B_{BBB}\lb \vec \ell_1, \vec \ell_2, \vec \ell_3 \rb= 0, \non
}
and in case of the cross-correlations, $B_{EBE}$ and $B_{BEE}$ are given by helicity permutations of $B^{(0,0,1)}_{222}$, 
while the rest of the angular bispectra, involving more then one $\gamma_B$ field are zero at tree-level.

\section{Results for one-loop power spectrum and tree-level bispectrum}
\label{sec:res}

In this section we show the angular power spectrum and bispectrum of intrinsic galaxy shapes, based on the results obtained in Paper I \cite{Vlah++:2019}.  
As we indicated in the previous section we show the results using the flat-sky approximation. The primary motivation for 
this lies in the realization that on the scales where one-loop terms are relevant, flat-sky is a good approximation to 
the all-sky result. We will address the return to the all-sky results in the subsequent paper.

Before we continue, we would like to point out the issue of the redshift dependence of the free bias parameters. 
Generally, the bias parameters are expected to evolve on a Hubble time scale, so that one can write
\eeq{
b_O(z) = b_O(\bar z) + b_O'(\bar z) (z-\bar z) +...
}
where $b_O'$ is of similar order as $b_O$. If the redshift distribution considered is not too wide, 
then one can approximately work with constant bias parameters (the correction is of order 
$(\Delta z)^2$ where $\Delta z$ is the width of the redshift distribution).
We will however not explore this issue further in this paper and we thus
assume that bias parameters are independent of redshift. 
  
\begin{figure}[t!]
\centering
\includegraphics[width=0.85\linewidth]{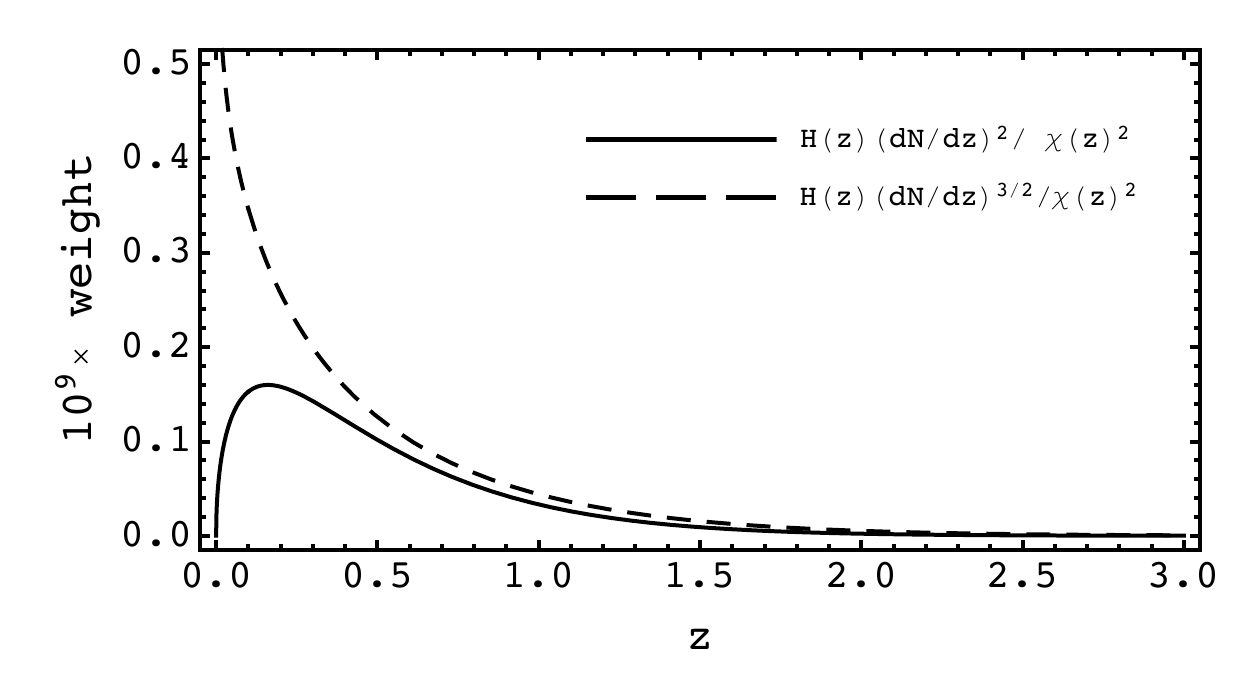}
\caption{
Power spectrum weight function (solid line) the compared to the square root of the bispectrum 
weight function (dashed line), based on Eq.~(\ref{eq:dNdz}). Both weights are multiplied by the factor $10^9$. The dashed curve diverges at low redshifts as $z^{3/2 \alpha - 2}$.}
\label{fig:ps_and_bias_waights}
\end{figure}

Furthermore, in order to show some concrete results we 
assume a simple weight function $W_g(\chi)$ that 
corresponds to the redshift distribution of galaxies with shapes expected for 
a next generation wide and deep survey. In other words,
\eeq{
  W_g(\chi)= \frac{dN}{dz}\frac{dz}{d\chi}
}
and where
\eeq{
\frac{dN}{dz}\propto z^\alpha \exp\left[-\left(\frac{z}{z_0}\right)^\beta\right],
\qquad {\rm and } \qquad
\frac{dz}{d\chi} = H(z),
\label{eq:dNdz}
}
e.g. \cite{Krause++:2015}, where $\alpha=1.24$, $\beta=1.01$ and $z_0=0.51$.
This specific example corresponds to the so-called ``gold sample'' of galaxies 
with shapes expected to be observed with the Rubin Observatory's Legacy Survey of Space and Time
\cite{Chang++:2013,Abell++:2009}. Note that the results shown here for shape correlations are for {\it intrinsic alignments alone}, i.e. without the contribution of the lensing signal. 
For simplicity, we assume that the same sample is used for both shapes and clustering, i.e. $W(\chi)\equiv W_{\rm n}(\chi) = W_g(\chi)$. This is generally not the case in galaxy surveys, which tend to focus on either spectroscopic or color-selected clustering samples to achieve better redshift accuracy. Nevertheless, we adopt this simple case here for illustration purposes. 
In Figure~\ref{fig:ps_and_bias_waights} we show 
how these weight functions contribute to the angular power spectrum and bispectrum.
We see that the weight function $W_g^2$ for the power spectrum peaks around the
redshift $z\approx 0.2$, while in case of bispectrum the weight function $W_g^3$
(the square of what is shown in the figure) diverges at low redshifts as $\propto z^{3 \alpha - 4}$.

\begin{figure}[ht]
\resizebox{\textwidth}{!}{
\centering
\includegraphics[width=\linewidth]{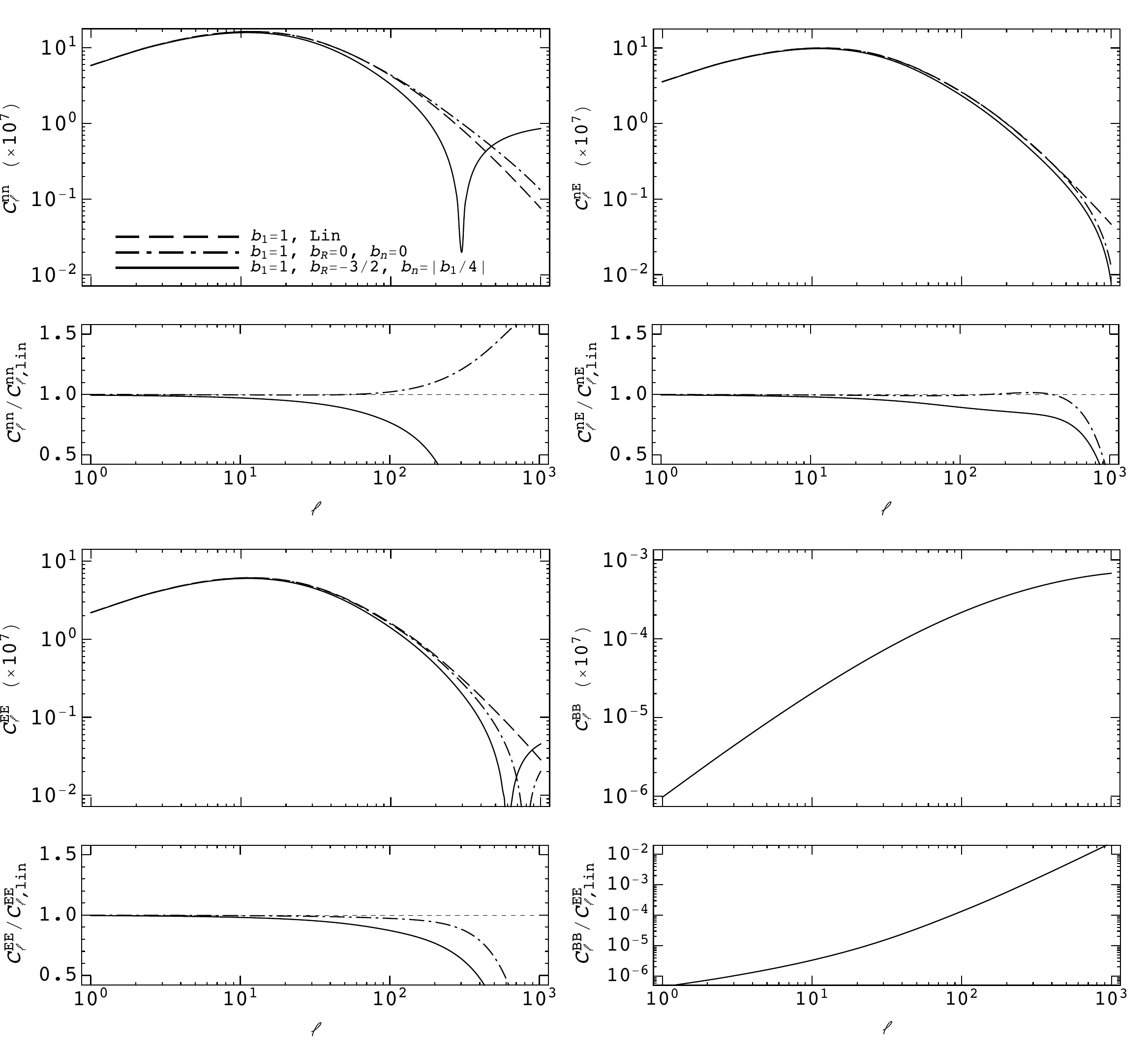}
}
\caption{The angular density and shape auto- and cross- power spectra. In each panel, 
upper plots show the angular power spectra, while lower plots show the ratio to the 
linear theory.
In the \textit{upper-left} panel, we show the galaxy number density-number density 
contribution $C^{{\rm nn}}_{\ell}$. Dashed lines are linear theory predictions containing 
only $b_1$ bias parameter, dot-dashed lines are nonlinear predictions again using only $b_1$ bias parameter,
while the solid lines show the full one-loop bias result. In the \textit{upper-right} panel, 
we show galaxy number density-$E$-mode angular power spectrum $C^{{\rm n} E}_{\ell}$. 
Dashed, dot-dashed, and solid lines again correspond to contributions as in the previous panel.
\textit{Lower-left} panel shows the $EE$ shape angular power spectrum, $C^{EE}_{\ell}$,
where again, three different lines are the same as in the previous panels.
Lastly, the \textit{lower-right} panel shows the $BB$ shape angular power spectrum $C^{BB}_{\ell}$.
Here the leading order contributions start at one-loop with the characteristic mode coupling dependence at large scales
that is highly suppressed relative to the $C_{EE,\rm lin}$ .
}
\label{fig:ang_powers_spectra}
\end{figure}

In Figure~\ref{fig:ang_powers_spectra} we show the various cross and auto angular 
power spectra for galaxy number density and $E$ and $B$-modes of galaxy shape. 
Below each sub-plot, we show the ratio between the terms obtained from the EFT 
one-loop prediction and the linear power spectrum. The exception is the bottom 
right case, since $C^{BB}_{\ell}$ is expected to be null at linear order. 
In this case, we normalize it to the $EE$ power spectrum. The first angular power 
spectrum (upper-left panel) is the galaxy number density-number density contribution 
$C^{\rm nn}_{\ell}$. We plot the EFT contributions to the angular power spectra 
as given in Section 5.1 of Paper I \cite{Vlah++:2019}. We show three curves corresponding 
to: the linear bias with linear matter power spectrum (dashed line),  
the linear bias with the one-loop matter power spectrum (dot-dashed line),  
and full EFT one-loop with non-linear biasing (solid line). In the last case we adopt the following values 
for the bias parameters: $b_1=1$, $b_{R_*}=-3/2$ and $b_n=b_1/4$ (all other bias coefficients), 
and we neglect the stochastic contributions. 
The adopted values for the bias parameters are chosen for visualization purposes only. In realistic cases, when the model 
is supposed to be contrasted with the data or simulations, these parameters would need to 
be fitted for. Similarly to this we show the contributions to the number density-shape angular 
power spectrum  $C^{{\rm n}E}_{\ell}$ (upper-right panel). Notice that the $b_1$ term in case of the shape spectra corresponds to the 
linear alignment model from \cite{Catelan++:2000}. 
From the ratio sub-plot shown below the $C^{\rm nn}_{\ell}$ prediction, and in general for all other 
correlations shown in Figure~\ref{fig:ang_powers_spectra}, we see that the EFT contributions are 
scale-dependent and have the highest impact at the smallest separations between galaxies 
(large $\ell$ values), as expected. In practice, the EFT expansion 
should only be trusted as long as the new terms are a fraction of the linear one.

The bottom panels of Figure~\ref{fig:ang_powers_spectra} show the intrinsic shape auto-correlations: 
$C^{EE}_{\ell}$ (left) and  $C^{BB}_{\ell}$ (right), using the same bias value prescription. Our predictions indicate that $C^{BB}_{\ell}$, which 
is null in linear alignment model predictions, is present at the mildly-nonlinear level. This is also 
in line with SPT predictions \cite{Blazek++:2017,Biagetti++:2020}. However, because $E$ and $B$-modes are 
connected by the same choice of biases, the bottom right panel of Figure~\ref{fig:ang_powers_spectra} 
can be interpreted as the typical order of magnitude of $B$-modes at mildly non-linear scales in 
comparison to the $E$-modes. We find these to be highly suppressed.  

\begin{figure}[t!]
\centering
\includegraphics[width=0.7\linewidth]{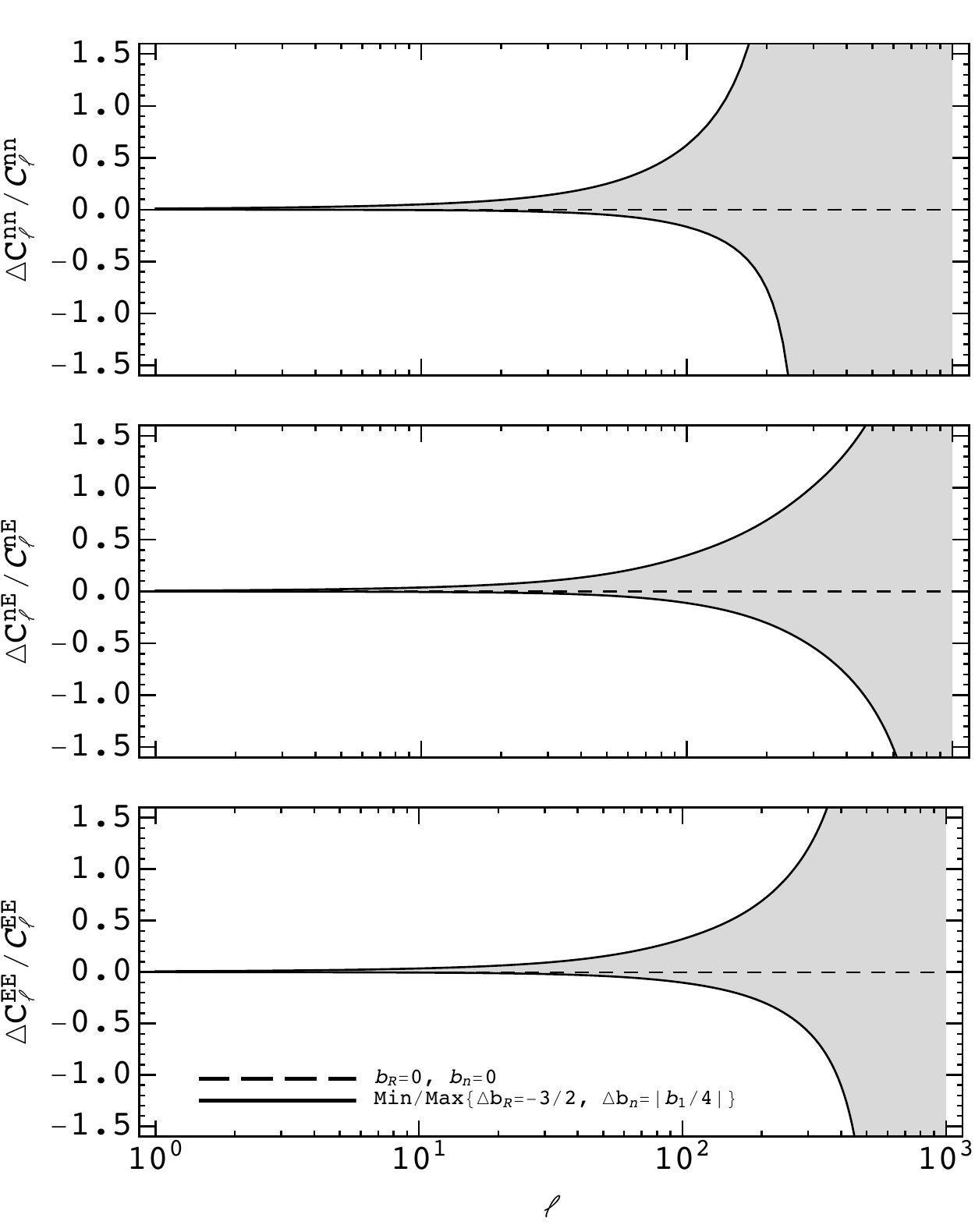}
\caption{Relative one-loop bias contributions compared to the dark matter one-loop
power spectra for $C^{{\rm nn}}_{\ell}$, $C^{{\rm n} E}_{\ell}$, $C^{EE}_{\ell}$. 
We again use $b_1=1$, $b_{R_*}= - 3/2$ and for the rest of the biases we use the values $|b_n| = b_1/4$. 
The sign of bias values $b_n$ are adjusted so that the grey area represents the band of potential bias one-loop contributions where $|b_n|<b_1/4$.
In all cases we neglect the stochastic bias contributions. Effectively this gives an estimate of the validity regime of linear alignment model, 
provided our bias coefficient choice is realistic.}
\label{fig:bias_var}
\end{figure}

In Figure~\ref{fig:bias_var}, we show a plausible range for fractional contribution of one-loop terms 
to each linear angular power spectrum, as a function of multipole and based on the variation of bias 
parameters. The variation of the bias parameters corresponds to the choice given in Figure 3
of Paper I \cite{Vlah++:2019}.
The three panels represent $C^{\rm nn}_{\ell}$, $C^{{\rm n} E}_{\ell}$ and $C^{EE}_{\ell}$. 
Note that we again neglect the stochastic bias contributions. The contributions are contained until scales of 
$l \sim 10^2$ and become comparable to the linear prediction for larger multipoles.

Since this paper is focused on intrinsic alignment contributions to
galaxy shape correlations, we have not included the contribution from
gravitational lensing here. Assuming that lensing and alignment effects
are additive, which holds to leading order, it is straightforward to
include gravitational lensing as described in Appendix \ref{app:lensing}.
The result for $C^{{\rm n}E}_{\ell}$ shown here then functions as a contaminant to galaxy-galaxy lensing,
while that for $C^{EE}_{\ell}$ is a contaminant to cosmic shear.
Note however that there are also important cross-contributions of intrinsic
alignments and lensing, which are included in the relations given in Appendix \ref{app:lensing}.

Next, in Figure~\ref{fig:ang_bispectra}, we show the number density auto- and 
number density, number density and shape cross-bispectra. As mentioned, for all bispectra 
cases we adopt the simplification for the window function $W_g = W_{\rm n}$,
thus all the weights are proportional to $W_g^3$.  We adopt the 
same bias values as in the case of power spectra above. The tree-level 3D bispectra needed 
for the computation of the angular bispectra here are explicitly given in Section 5.2 
of Paper I \cite{Vlah++:2019}. All panels are normalized by the maximal value attained over the configurations shown.
In the upper panels we show the contributions for the number density auto-bispectrum
$B_{\rm {nnn}}$ for configurations with maximal wavenumber $\ell_1 = 30$ (left panel)
and $\ell_1 = 150$ (right panel). In both cases we see that the equilateral contributions 
are suppressed compared to the $\ell_2+\ell_3 \approx \ell_1$ diagonal (flattened triangles), where 
the amplitude the bispectrum amplitudes reach their maximal values.
Middle panel shows the number density, number density and $E$-mode 
cross-bispectrum $B_{{\rm nn} E}$  for the same maximal wavenumber $\ell_1$ choices. 
We notice the similar relative configuration distribution as in the case of $B_{\rm {nnn}}$, 
with the equilateral contributions suppressed relative to the $\ell_2+\ell_3 \approx \ell_1$ diagonal. 
The property that the bispectrum amplitudes peak for the flattened triangles, where $\ell_2+\ell_3 \approx \ell_1$,
is inherited from the similar behavior of the 3D bispectrum of matter
and biased tracers (see the right panels of Fig.~10 in \cite{Desjacques++:2016}).
 
Similar to the $E$-mode, in the lower panels we show the number density, 
number density and $B$-mode cross-bispectrum $B_{{\rm nn}B}$ . For the same 
choice of bias parameters as before, the maximal amplitude of this bispectrum is 
suppressed by an order of magnitude relative to the $B_{\rm {nnn}}$ and $B_{{\rm nn}E}$. 
Additionally, relative to this maximal amplitude configuration, the rest of the configurations 
are further suppressed more strongly than in the $B_{\rm {nnn}}$ and $B_{{\rm nn}E}$ cases. 
In other words, in the $B_{{\rm nn}B}$ cross-bispectrum, most of the relevant 
contributions are roughly centered around the maximal amplitude configuration
and they are again clustered close to the flattened triangle shapes, i.e. 
$\ell_2+\ell_3 \approx \ell_1$ is diagonal. The latter feature can again be 
explained by the shape of the 3D bispectra.

Lastly, let us comment on the fact that the maximal $B_{{\rm nn}B}$ 
contribution is only one order of magnitude suppressed relative to $B_{{\rm nn}E}$. 
This is in stark difference compared to the power spectrum case, 
where the only channel of comparison was $C^{BB}_{\ell}$ compared to $C^{EE}_{\ell}$. 
There, the large suppression of $C^{BB}_{\ell}$ on large scales was caused by the fact that 
the leading order contribution was a one-loop result, versus the linear one for the $C^{EE}_{\ell}$. 
In the case of $B_{{\rm nn}E}$ and  $B_{{\rm nn}B}$, this is no longer so, and 
both of these bispectra have non-vanishing tree-level contributions,
so that the $B$-mode contribution is relatively much less suppressed.

\begin{figure}[t!]
\centering
\includegraphics[width=0.98\linewidth]{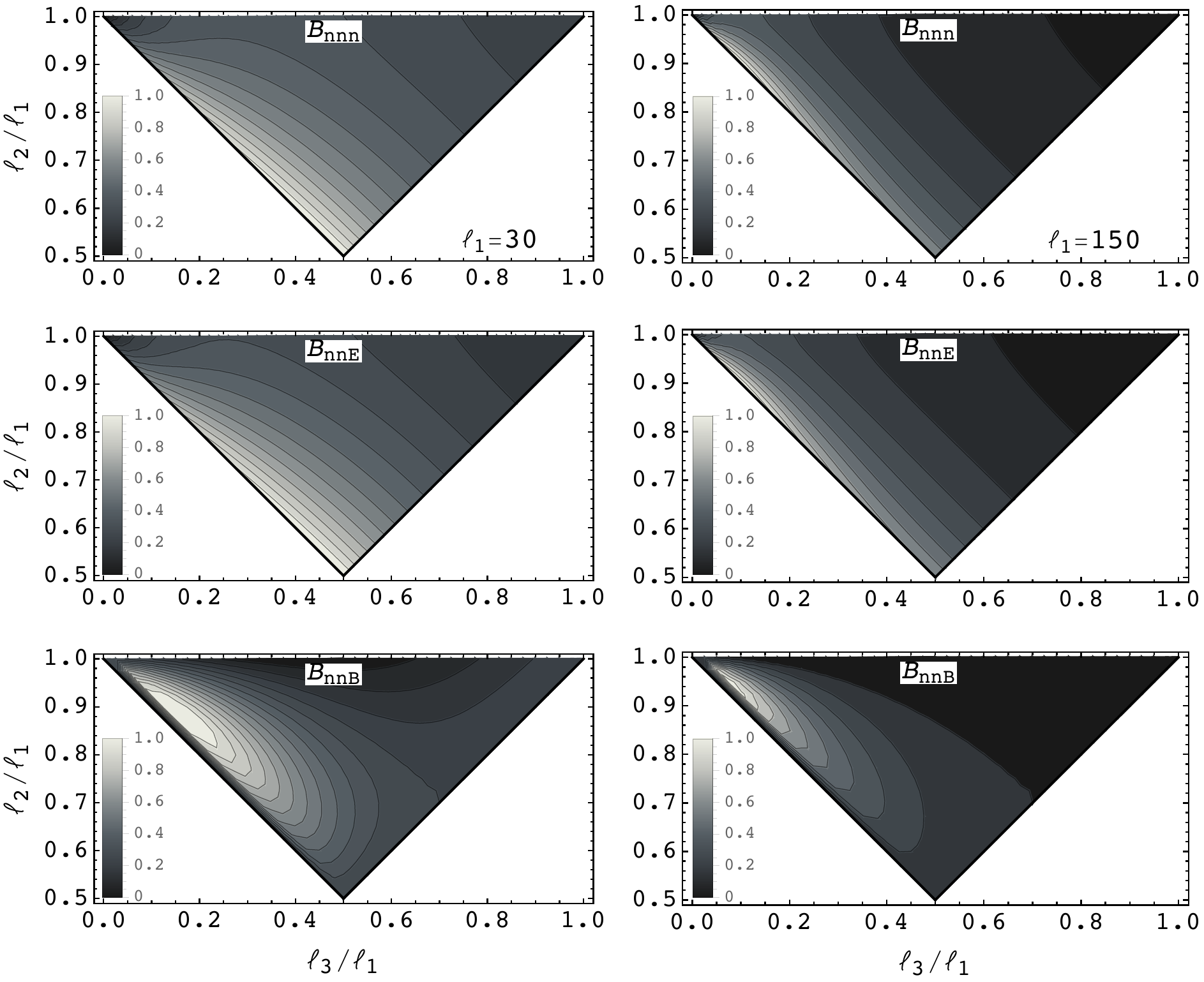}
\caption{The angular number density and shape auto- and cross-bispectrum. 
Left panels show bispectra with maximal wavenumber $\ell_1 = 30$
while on the right panels it is $\ell_1 = 150$.
Upper panels show the number density auto-bispectrum $B_{\rm {nnn}}$, 
while the middle and lower panels show the cross-bispectrum of two number 
density and one shape fields in the $B_{{\rm nn}E}$ and $B_{{\rm nn}B}$ cross-bispectra 
configurations, respectively. All bispecra are normalized to its maximal amplitude configuration value. 
The maximal amplitudes are, typically, an order of magnitude smaller in $B_{{\rm nn}B}$ 
compared to $B_{{\rm nn}E}$, for both  $\ell_1 = 30$ and $\ell_1 = 150$ cases. 
Values of the bias parameters are the same as in the case of power spectra. 
}
\label{fig:ang_bispectra}
\end{figure}

\section{Conclusions}
\label{sec:concl}

In Paper I \cite{Vlah++:2019}, we had presented an EFT expansion 
of galaxy intrinsic shapes, sizes and number density tracers that allows 
one to perform a complete prediction of three-dimensional Fourier-space 
statistics, i.e. power spectra up to the one-loop order and tree-level bispectra. 
In this work, we developed the projection formalism that relates these 
3D correlators, which describe the physical dynamics of the observables, 
with the projected quantities that live on the 2D plane of the sky. 
We adopted the \textit{flat-sky} approximation for computing the angular 
power and bispectra. This provides an accurate description of angular power spectrum 
on scales where our one-loop corrections apply. Furthermore, our results, 
which establish the connection between components of the 3D statistics 
and projected 2D  statistics, are applicable to more general 3D correlators, 
and not just the ones predicted by the EFT of Paper I. Indeed, these projections 
are performed independently from the methods by which the 3D statistics 
is obtained and instead rely entirely on the underlying statistical symmetries 
(homogeneity, isotropy and parity invariance). 

We presented results for the angular two- and three-point statistics
of galaxy counts and shapes. The results are presented using three different bases; the 
helicity basis, which is the natural basis for applying the above-mentioned 
symmetry constraints; the electric-magnetic component basis, which is the
commonly used basis in the field and in which we present our final results; 
and lastly, the Pauli basis that is also frequently used in the lensing community. 
The development of this formalism allows one to identify the forms of 
degeneracies taking place when projecting into the angular power spectra 
and bispectra, as well as which contributions survive at the tree-level bispectrum 
and one-loop power spectrum. In case of the power spectrum, the situation 
is relatively simple, and all number density and $E$-mode auto- and 
cross-correlations have contributions already at linear order. In contrast, the 
cross-correlations of these with the $B$-mode vanish to all orders as 
a consequence of the statistical parity invariance (in addition to the homogeneity
and isotropy). The remaining $B$-mode auto correlation gets its leading order 
contribution from the one-loop mode-coupling terms. It is for this reason several
orders of magnitude suppressed compared to the number density and $E$-mode 
correlators (on the scales of our interest) which, as mentioned, benefit in that 
respect from the leading linear order contribution. In the case of bispectra, 
the statistical symmetry considerations are less constraining allowing for all 
possible number density, $E$-mode and $B$-mode auto- and cross-correlations. 
However, many of these do not contribute in the tree-level PT, appearing 
only at higher-loop orders. This is typically the case in bispectra that correlate 
more than one $B$-mode. An interesting observation in case of B-mode 
contributions to the bispectrum is that it no longer is as strongly suppressed (as was 
the case in the power spectrum) compared to the other correlations. The reason 
for this is that all of the considered cross-correlations ($B_{{\rm nnn}}$, 
$B_{{\rm nn}E}$ and $B_{{\rm nn}B}$) have contributions starting at the 
tree-level in PT. 

Assuming plausible bias parameter values, we made estimates 
of these correlators using the EFT results of Paper I \cite{Vlah++:2019},
and we present several numerical results for the statistics discussed above. 
The current study is only qualitative and for more robust quantitative performance 
assessment, comparisons to simulations would be needed. On the other hand, our formalism 
can be readily applied to ongoing weak gravitational lensing surveys of 
the large-scale structure to model position-shape and shape-shape 
correlations, as well as three-point statistics, in a consistent and complete 
manner up to quasi-linear scales. Moreover, we emphasize that its potential 
applications are even more general: it can be applied to any study of 
tensorial fields on the sky. In the future, we plan to present analogous \textit{all-sky} 
derivations that can improve the accuracy of our predictions at low $\ell$ values, 
as well as predictions for configuration space correlation functions in this regime. 

\acknowledgments

NEC was supported by a Royal Astronomical Society Research Fellowship for part of this project. 
This work is also part of the Delta ITP consortium, a program of the Netherlands Organisation for Scientific Research (NWO) that is funded by the Dutch Ministry of Education, Culture and Science (OCW).
FS is supported by Starting Grant (ERC-2015-STG 678652) ``GrInflaGal'' of the European Research Council.
ZV is supported by the Kavli Foundation.

\appendix

\section{Gravitational lensing}
\label{app:lensing}

We now provide expressions that include the gravitational lensing
contribution to shapes, under the approximations stated after
Eq. (\ref{eq:g_projected}):
first, we assume that lensing simply adds to the intrinsic shape; second,
we neglect all post-Born, reduced-shear and lensing-bias corrections, so
that $\gamma_{G,ij}$ can be written as the second derivative of a lensing
potential, which in turn is proportional to the gravitational potential.
This means that $\gamma_G$ only receives helicity-0 contributions.

In the polarization basis, defined in \refeq{polarization_gamma}, the
angular power spectra including lensing can then be written as
\eq{
  C_{{\rm n} E}(\ell) &= \int \frac{d\chi}{\chi^2} \bigg[  \frac{1}{2}  W_{\rm n}(\chi)W_g(\chi) N_0 P^{(0)}_{02} ( \ell / \chi)
    +  W_{\rm n}(\chi)W_G(\chi) P_{\rm nm} ( \ell / \chi) \bigg] ,
  \label{eq:Cl_flatsky_lensing}\\
  C_{EE}(\ell) &= \int \frac{d\chi}{\chi^2}
  \bigg[  \frac{1}{8} W_g^2(\chi) \lb 2 N_0^2 P^{(0)}_{22} ( \ell / \chi) + N_2^2 P^{(2)}_{22} ( \ell / \chi) \rb    \non\\
    & \hspace*{1.6cm} + W_g(\chi) W_G(\chi) N_0 P^{(0)}_{2{\rm m}} ( \ell / \chi)
    + W_G^2(\chi) P_{\rm mm} ( \ell / \chi)
    \bigg] , \non\\
C_{BB}(\ell) &= \int d\chi  \frac{W_g^2(\chi)}{\chi^2} N_1^2 P^{(1)}_{22} ( \ell / \chi) .  \non
  }
  Here,
  \be
W_G(\chi) = \frac{3}{2} \Omega_{\rm m}H_0^2 \frac{\chi}{a}\int_\chi^{\infty}d\chi'\, W_g(\chi') \frac{\chi'-\chi}{\chi}
\ee
is the lensing kernel in the absence of curvature. 
$P_{{\rm nm}}(k,z)$ is the cross-power spectrum between matter 
density and number counts n (which can be calculated up to 1-loop order using the EFT of 
biased tracers), while $P_{\rm mm}(k,z)$ is the nonlinear matter power spectrum. 
$P^{(0)}_{2{\rm m}}(k,z)$ is the cross-correlation between matter and the helicity-2 component 
of the intrinsic shape. This can be obtained in the EFT of shapes from $P^{(0)}_{02}(k,z)$ 
by setting all the number-count bias parameters to zero, except for $b_1$ which is set to 1.
For explicit expressions for the EFT power spectra, we refer the reader to \cite{Vlah++:2019}.
We defer the inclusion of the lensing contribution to bispectra statistics \cite[e.g.][]{Bartelmann+:1999,Takada+:2003} to future work. 

\section{Angular power spectrum}
\label{app:flat_sky_spectrum}

Let us assume we have a real tensor field $X_{ij}(\vec r)$ such that we can identify the 
trace part as the simple scalar overdensity $\df(\vec r)$, while the trace-free part 
can be decomposed into the helicity two components $\gamma_{\pm2}(\vec r)$. 
In other words, we can write 
\eeq{
X_{ij}(\vec r, z) = \df(\vec r, z)  \vec M^{(0)}_{ij}(\vhat r) + \gamma_{+2}( \vec r, z)  \vec M^{(+2)}_{ij}(\vhat r) + \gamma_{-2}( \vec r, z)  \vec M^{(-2)}_{ij}(\vhat r),
\label{eq:Xij_decomposition}
}
where we have introduced (only for this section) $\vec M^{(0)}_{ij}(\vhat r) = \dK_{ij}/{\sqrt 3}$.
It is convenient to use this field in order to compute the relevant statistics. 
Moreover, we assume that the power spectrum of this field is of the form 
given in \refeq{decomp_ps_main}, i.e. we have
\eeq{
\la  X_{ij}   ( \vec k_1, z_1)   X_{kl}  ( \vec k_2, z_2) \ra = (2\pi)^3 \dD(\vec k_1 + \vec k_2) P_{ijkl}(\vec k, z_1,z_2).
}

We are interested in angular statistics on the sky. Here, we focus on the flat-sky approximation results and thus we can introduce the field (see Figure \ref{fig:flat_sky} for coordinate setup)
\eeq{
 \hat X_{s}(\vec \ell)   = \int d^2\vec \theta ~ \hat X_s (\vec \theta) e^{si(\phi_\theta - \phi_\ell)} e^{i \vec \ell . \vec \theta},
 \label{eq:X_ell_flat}
}
in analogy to \refeq{gamma_2D_FT}, and where
\eeq{
\hat X_s (\vec \theta) = \vec M^{(s)*}_{ij} (\vhat n) \int d\chi ~ W(\chi) X_{ij}\big( \chi \vhat n , \chi \vec \theta, z[\chi] \big).
 \label{eq:X_flat_FT}
}
The flat-sky projected two-point function can be expressed as 
\eeq{
\la  X_{ij} \big( \chi_1 \vhat n , \chi_1 \vec \theta_1 \big) X_{kl} \big( \chi_2 \vhat n , \chi_2 \vec \theta_2 \big) \ra 
= \int \frac{d^3 k}{(2\pi^3)} ~ P_{ijkl}(\vec k) e^{ - i (\chi_2 - \chi_1) \vhat n . \vec k - i ( \chi_2 \vec \theta_2 - \chi_1 \vec \theta_1 ).\vec k} ,
}
where
\eq{
\la X_{s_1} (\vec \theta_1)  X^*_{s_2} (\vec \theta_2) \ra 
&=  \vec M^{(s_1)*}_{ij} (\vhat n)  \vec M^{(s_2)}_{kl} (\vhat n) \int d\chi_1 d\chi_2  ~ W(\chi_1) W(\chi_2) \\
&\hspace{2.5cm} \times  \int \frac{d^3 k}{(2\pi^3)} ~ P_{ijkl}(\vec k) e^{ - i (\chi_2 - \chi_1) \vhat n . \vec k - i ( \chi_2 \vec \theta_2 - \chi_1 \vec \theta_1 ).\vec k}.  \non
}
It is convenient to setup the coordinate frame so that $k_\pp =  \vhat n . \vec k$ and $\vec k_\perp = \vec k - k_\pp \vhat n$, and 
using the variables $\chi_{2,1} = \chi \pm \frac{1}{2}\Delta \chi$. This gives us
\eq{
&\int d\chi_1 d\chi_2  ~ W(\chi_1) W(\chi_2) \int \frac{d^3 k}{(2\pi^3)} ~ P_{ijkl}(\vec k) e^{ - i (\chi_2 - \chi_1) \vhat n . \vec k - i ( \chi_2 \vec \theta_2 - \chi_1 \vec \theta_1 ).\vec k} \\
&= \int d\chi d\Delta\chi  ~ W\lb \chi+  \tfrac{1}{2} \Delta \chi\rb W\lb \chi -  \tfrac{1}{2} \Delta \chi\rb \non\\
&\hspace{3cm} \times \int \frac{d k_\pp d^2\vec k_\perp }{(2\pi)^3} ~ P_{ijkl}( k_\pp \vhat n + \vec k_\perp)  
e^{- i k_\pp \Delta \chi  - i \chi ( \vec \theta_2 - \vec \theta_1 ).\vec k_\perp - i  \frac{1}{2}\Delta \chi \lb \vec \theta_2 + \vec \theta_1 \rb .\vec k_\perp } \non
}
To further simplify the calculation note that the leading wavenumber $k$ modes that can contribute to the integral 
are those ones with small $k_\pp =  \vhat n . \vec k$, so that $k_\pp \ll 1/(\chi\theta)$. Since we work 
in the small-angle approximation, we have $1/(\chi\theta) \gg 1/\chi$ and thus modes with longitudinal 
wavenumber $k_\pp$ much larger than $1/\chi$ do not give rise to angular correlations because of cancellations 
along the line of sight. Only modes with $k_\pp$ similar to $1/\chi$ lead to angular correlations. 
The relevant transverse wavenumbers $1/(\chi \theta)$ are thus much larger than the relevant 
longitudinal wavenumbers, and we can set the argument of the spherical components of the tensor power spectrum to 
\eeq{
P^{(m)}_{\ell \ell'} (k) = P^{(m)}_{\ell \ell'} (k) \lb \sqrt{ k_\pp + k_\perp } \rb \approx P^{(m)}_{\ell \ell'} (k_\perp),
}
from which it follows that 
\eeq{
P_{ijkl}(\vec k) = P_{ijkl}(k_\pp \vhat n + \vec k_\perp) \approx P_{ijkl}(\vec k_\perp) .
}
We can now integrate over  $k_\pp$ which gives us the delta function that sets $\Delta \chi = 0$ 
and 
\eq{
&\int d\chi_1 d\chi_2  ~ W(\chi_1) W(\chi_2) \int \frac{d^3 k}{(2\pi^3)} ~ P_{ijkl}(\vec k) e^{ - i (\chi_2 - \chi_1) \vhat n . \vec k - i ( \chi_2 \vec \theta_2 - \chi_1 \vec \theta_1 ).\vec k} \\
&\hspace{5cm}\approx \int d\chi  ~ [W\lb \chi \rb]^2 \int \frac{d^2\vec k_\perp }{(2\pi)^2} P_{ijkl}(\vec k_\perp)  e^{- i \chi ( \vec \theta_2 - \vec \theta_1).\vec k_\perp} . \non
}
This allows us to compute the flat-sky projected angular power spectrum. 
Using the definition in \refeq{X_ell_flat} and the expression above we have
\eq{
\la \hat X_{s_1}(\vec \ell_1)  \hat X^*_{s_2}(\vec \ell_2)  \ra
&= 
 \int d^2\vec \theta_1 d^2\vec \theta_2 \la \hat X_{s_1} (\vec \theta_1)  \hat X^*_{s_2} (\vec \theta_2) \ra  
 e^{s_1i(\phi_{\theta_1} - \phi_{\ell_1}) - s_2i(\phi_{\theta_2} - \phi_{\ell_2})} e^{i \vec \ell_1 . \vec \theta_1 - i \vec \ell_2 . \vec \theta_2} \\
&\approx  (2\pi)^2 \dD \lb \vec \ell_1 - \vec \ell_2 \rb 
\int d\chi  \frac{[W\lb \chi \rb]^2}{\chi^2} \vec {\widetilde M}^{(s_1)*}_{ij} (\vhat n)  \vec {\widetilde M}^{(s_2)}_{kl} (\vhat n) P_{ijkl}( \vec \ell_2/\chi ). \non
}
We also used the fact that 
\eq{
 \int d^2\vec \theta ~  \vec M^{(s)}_{ij} (\vhat n)
  e^{ - si(\phi_\theta - \phi_{\ell})} e^{- i \vec \ell . \vec \theta} 
= (2\pi)^2 \dD \lb \vec \ell \rb \vec {\widetilde M}^{(s)}_{ij} (\vhat n).
\label{eq:Mij_FT}
}
Given the decomposition of $P_{ijkl}$ given in \refeq{decomp_ps_main} we have 
\eq{
\vec {\widetilde M}^{(s_1)*}_{ij} (\vhat n)  \vec {\widetilde M}^{(s_2)}_{kl} (\vhat n) P_{ijkl}( \vec \ell_2/\chi )
&= \frac{1}{|s_1|!|s_2|!} \bigg( N^2_0 P^{(0)}_{|s_1||s_2|}(\ell_2/\chi) \\
&\hspace{1.5cm} + s_1 s_2 N^2_1 P^{(1)}_{|s_1||s_2|}(\ell_2/\chi) + \frac{(s_1 s_2)^2}{32} N^2_2 P^{(2)}_{|s_1||s_2|}(\ell_2/\chi) \bigg). \non
}
Using the definition of flat-sky angular power spectrum
\eeq{
\la  \hat X_{s_1}(\vec \ell_1)  \hat X_{s_2}(\vec \ell_2)  \ra
= (2\pi)^2 \dD \lb \vec \ell_1 + \vec \ell_2  \rb C_{s_1 s_2} \lb \ell_1 \rb,
}
we arrive at the final expression:
\eq{
\label{eq:flat_sky_angular_Cl}
C_{s_1 s_2} (\ell) = \frac{1}{|s_1|!|s_2|!} \int d\chi  \frac{[W\lb \chi \rb]^2}{\chi^2}  \bigg( N^2_0 P^{(0)}_{|s_1||s_2|}(\ell/\chi) &+ s_1 s_2 N^2_1 P^{(1)}_{|s_1||s_2|}(\ell/\chi)  \\
&+ \frac{(s_1 s_2)^2}{32} N^2_2 P^{(2)}_{|s_1||s_2|}(\ell/\chi) \bigg), \non
}
where the normalization constants $N_{0,1,2}=\left\{ \sqrt{3/2}, \sqrt{1/2}, 1\right\}$ 
originate from the basis $\vec Y^{(m)}_{ij}$, introduced in \refeq{Y_basis}.
Since all the components of the $C_{s_1 s_2}$ angular power spectra are real, we have
\eeq{
\la \hat X_{s_1}(\vec \ell_1)  \hat X^*_{s_2}(\vec \ell_2)  \ra = \la \hat X^*_{s_1}(\vec \ell_1)  \hat X_{s_2}(\vec \ell_2)  \ra .
}

\section{Angular bispectrum}
\label{app:flat_sky_bispectrum}

To obtain the flat-sky expressions for the angular bispectrum 
we combine \refeqs{X_ell_flat}{X_flat_FT} 
to get 
\eq{
 \hat X_{s}(\vec \ell)  &=
  \int d\chi ~ W(\chi) 
  \int d^2\vec \theta ~ \vec M^{(s)*}_{ij} (\vhat n) X_{ij}\big( \chi \vhat n , \chi \vec \theta, z[\chi] \big)
 e^{si(\phi_\theta - \phi_\ell)} e^{i \vec \ell . \vec \theta}, \\
&=
  \int d\chi ~ W(\chi) 
  \int d^2\vec \theta ~ \vec M^{(s)*}_{ij} (\vhat n)  e^{si(\phi_\theta - \phi_\ell)} e^{i \vec \ell . \vec \theta}
  \int \frac{dk_\pp d^2k_\perp}{(2\pi)^3} X_{ij}\big( \vec k, z[\chi] \big) ~e^{ - i \chi k_\pp - i \chi \vec k_\perp . \vec \theta} \non\\
&=
  \int d\chi ~ W(\chi) 
  \int \frac{dk_\pp d^2k_\perp}{2\pi} ~e^{ - i \chi k_\pp}
  \dD \lb \vec \ell - \chi \vec k_\perp \rb \vec {\widetilde M}^{(s)*}_{ij} (\vhat n)
  X_{ij}\big( k_\pp \vhat n, \vec k_\perp, z[\chi] \big) \non\\
&=
  \int d\chi ~ \frac{W(\chi)}{\chi^2} 
  \int \frac{dk_\pp }{2\pi} ~e^{ - i \chi k_\pp} \vec {\widetilde M}^{(s)*}_{ij} (\vhat n)
  X_{ij}\big( k_\pp \vhat n, \vec \ell/\chi, z[\chi] \big). \non
}
Using the definition of the 3D bispectrum given in \refeq{Bispectrum_3D}, and the same set of
approximations given in appendix \ref{app:flat_sky_spectrum} (e.g. $k_\pp \ll k_\perp$), we have
\eq{
 \int \frac{dk_{\pp,123}}{(2\pi)^3}
& ~e^{ - i \sum_l \chi_{l} k_{\pp,l}}
\la X_{ij}\big( k_{\pp,1}, \vec \ell_1/\chi_1, z_1 \big) X_{nm}\big( k_{\pp,2}, \vec \ell_2/\chi_2, z_2 \big)  X_{rs}\big( k_{\pp,3}, \vec \ell_3/\chi_3, z_3 \big) \ra \\
&= \chi^2_3 \dD(\chi_1-\chi_3)\dD(\chi_2-\chi_3) (2\pi)^2 \dD \lb \vec \ell_1 + \vec \ell_2 +\vec \ell_3  \rb
B_{ijnmrs} \lb \vec \ell_1/\chi_1,  \vec \ell_2/\chi_2,  \vec \ell_3/\chi_3 \rb. \non
}
Using the definition of flat-sky angular bispectrum
\eeq{
\la  \hat X_{s_1}(\vec \ell_1)  \hat X_{s_2}(\vec \ell_2)   \hat X_{s_3}(\vec \ell_3)  \ra
= (2\pi)^2 \dD \lb \vec \ell_1 + \vec \ell_2 +\vec \ell_3  \rb B_{s_1 s_2 s_3} \lb \vec \ell_1, \vec \ell_2, \vec \ell_3 \rb,
}
we have
\eq{
B_{s_1 s_2 s_3} \lb \vec \ell_1, \vec \ell_2, \vec \ell_3 \rb = 
\int d\chi ~ \frac{[W(\chi)]^3}{\chi^4} 
\vec {\widetilde M}^{(s_1)*}_{ij} (\vhat n)
\vec {\widetilde M}^{(s_2)*}_{nm} (\vhat n)
\vec {\widetilde M}^{(s_3)*}_{rs} (\vhat n)
B_{ijnmrs} \lb \tilde{\vec \ell}_1, \tilde{\vec \ell}_2, \tilde{\vec \ell}_3 \rb,
}
where $\tilde{\vec \ell}_i = \vec \ell_i/\chi$.
Under the assumptions that $k_\pp \ll k_\perp$ we can evaluate the basis $\vec Y^{(q)}$
from \refeq{Y_basis} by setting the $\vhat k$ so that it lays on one of the $z$ planes (see figure \ref{fig:flat_sky}).
We thus have
\eeq{
  \vec {\widetilde M}^{(s)*}_{ij} (\vhat n) \vec {Y}^{(q)}_{ij} (\vhat n) = i^{s-2q} c_{s,q},
  \label{eq:c5}
}
where $c_{s,q}$ are the following coefficients
\begin{center}
\begin{tabular}{ c | c c c c c } 
 $c_{s,q}$  & $-2$ & $-1$ & $0$ & $+1$ & $+2$ \\ \hline 
 $+2$ & $1/4$ & $i/2$ & $-1/2 \sqrt{3/2}$ & $-i/2$ & $1/4$ \\ 
 $-2$ & $1/4$ & $-i/2$ & $-1/2 \sqrt{3/2}$ & $i/2$ & $1/4$ \\ 
\end{tabular}
\end{center}

Using these results and \refeq{decomp_ps_main}, one can also recover the angular 
power spectrum results given in \refeq{flat_sky_angular_Cl_main} and 
derived explicitly in appendix \ref{app:flat_sky_spectrum}.
However, in this section we focus on the angular bispectrum results. 
Using the 3D bispectrum decompositions as given in section 
\refeq{bis_decomp_0} we have
\eq{
\label{eq:angular_bispectrum_flat_sky_app}
B_{{\rm n}{\rm n}{\rm n}} \lb \vec \ell_1, \vec \ell_2, \vec \ell_3 \rb & = \int d\chi ~ \frac{[W_{\rm n}(\chi)]^3}{\chi^4}  B^{(0)}_{000} \lb \tilde{\vec \ell}_1, \tilde{\vec \ell}_2, \tilde{\vec \ell}_3 \rb, \\
B_{{\rm n}{\rm n} s_3} \lb \vec \ell_1, \vec \ell_2, \vec \ell_3 \rb & = \frac{1}{2} \int d\chi ~ \frac{[W_{\rm n}(\chi)]^2 W_g(\chi)}{\chi^4}
\bigg[ N_0 B^{(0)}_{002} - i s_3 \sqrt{2} N_1 B^{(1)}_{002}  - N_2 B^{(2)}_{002} \bigg] \lb \tilde{\vec \ell}_1, \tilde{\vec \ell}_2, \tilde{\vec \ell}_3 \rb , \non\\
B_{{\rm n} s_2 s_3} \lb \vec \ell_1, \vec \ell_2, \vec \ell_3 \rb & =  \frac{1}{8} \int d\chi ~ \frac{W_{\rm n}(\chi) [W_g(\chi)]^2}{\chi^4}
\bigg[
2 N_0^2  B^{(0,0)}_{022} 
+ 2 s_2 s_3  N_1^2  \lb B^{(-1,1)}_{022} - B^{(1,1)}_{022} \rb \non\\
&\hspace{3.0cm} - i 2 \sqrt{2} N_0 N_1 \lb s_3 B^{(0,1)}_{022} + s_2 B^{(1,0)}_{022} \rb \non\\
&\hspace{3.0cm} + i \sqrt{2} N_1 N_2 \lb s_2 B^{(1,2)}_{022} + s_3 B^{(2,1)}_{022} + s_3 B^{(-2,1)}_{022} + s_2 B^{(1,-2)}_{022} \rb \non\\
&\hspace{3.0cm} - 2 N_0 N_2  B^{(\{0,2\})}_{022} 
+ N_2^2  \lb B^{(-2,2)}_{022} + B^{(2,2)}_{022} \rb 
\bigg]  \lb \tilde{\vec \ell}_1, \tilde{\vec \ell}_2, \tilde{\vec \ell}_3 \rb , \non\\
B_{s_1 s_2 s_3} \lb \vec \ell_1, \vec \ell_2, \vec \ell_3 \rb & =  \frac{1}{32}  \int d\chi ~ \frac{[W_g(\chi)]^3}{\chi^4}
\bigg[
4 N_0^3  B^{(0,0,0)}_{222}  - 4 N_0^2 N_2 B^{(\{0,0,2\})}_{222} \non\\
&\hspace{0.5cm} - N_2^3 \lb B^{(\{-2,2,2\})}_{222} + B^{(2,2,2)}_{222} \rb \non\\
&\hspace{0.5cm} + 4 N_0 N_1^2 \lb s_2 s_3 \lb B^{(0,-1,1)}_{222} - B^{(0,1,1)}_{222} \rb + 2~{\rm cyc. }  \rb \non\\
&\hspace{0.5cm} + 2 N_0 N_2^2 \lb B^{(\{0,-2,2\})}_{222} + B^{(\{0,2,2\})}_{222} \rb \non\\
&\hspace{0.5cm} + 2 N_1^2 N_2 \lb s_1 s_2 \lb B^{(1,1,2)}_{222} + B^{(1,1,-2)}_{222} - B^{(\{-1,1\},2)}_{222} \rb  + 2~{\rm cyc. } \rb \non\\
&\hspace{0.5cm} - i 4 \sqrt{2} N_0^2 N_1 \lb s_3 B^{(0,0,1)}_{222} + 2~{\rm cyc. }  \rb \non\\
&\hspace{0.5cm} - i 2 \sqrt{2} s_1s_2s_3 N_1^3 \lb B^{(\{-1,1,1\})}_{222} - B^{(1,1,1)}_{222} \rb \non\\
&\hspace{0.5cm} - i \sqrt{2} N_1 N_2^2 \lb s_3 \lb B^{(\{-2,2\},1)}_{222} + B^{(2,2,1)}_{222} - B^{(2,2,-1)}_{222} \rb + 2~{\rm cyc. }  \rb \non\\
&\hspace{0.5cm} + i 2 \sqrt{2} N_1 N_2 N_3 \lb s_3 \lb B^{(\{0,-2\},1)}_{222} + B^{(\{0,2\},1)}_{222} \rb + 2~{\rm cyc. }  \rb \bigg]   \lb \tilde{\vec \ell}_1, \tilde{\vec \ell}_2, \tilde{\vec \ell}_3 \rb. \non
}
Where $(\{a,b,c\})$ notation is used and we understand that the summation over all the nontrivial index permutations should be performed.
Above we also used the bispectrum parity relation given in \refeq{bis_parity} to reduce the number of independent terms. 
Each of the cross angular bispectra has contributions from different helicities, which give rise to the terms in the integrand. 
Depending on the number of helicity two fields in the angular bispectrum number of terms in the integrand varies from three for the 
density-density-shear (002) bispectrum, six for the density-shear-shear (022) bispectrum and ten for the shear-shear-shear (222). 
These multiplicities correspond to the number of combinations with repetition for the helicity of the shear field.

In Paper I \cite{Vlah++:2019} we gave the explicit bispectrum results for biased tracers up to the tree-level (leading order) PT.
In that case only, few of the terms above survive given that leading order fields carry only helicity zero contribution.
In other words, only single non-zero helicity terms survive and we have 
\eq{
\label{eq:LO_bis}
B^{\rm LO}_{{\rm n}{\rm n}{\rm n}} \lb \vec \ell_1, \vec \ell_2, \vec \ell_3 \rb & = \int d\chi ~ \frac{[W_{\rm n}(\chi)]^3}{\chi^4}  B^{(0)}_{000} \lb \tilde{\vec \ell}_1, \tilde{\vec \ell}_2, \tilde{\vec \ell}_3 \rb, \\
B^{\rm LO}_{{\rm n}{\rm n} s_3} \lb \vec \ell_1, \vec \ell_2, \vec \ell_3 \rb& = \frac{1}{2} \int d\chi ~ \frac{[W_{\rm n}(\chi)]^2 W_g(\chi)}{\chi^4}
\bigg[ N_0 B^{(0)}_{002} - i s_3 \sqrt{2} N_1 B^{(1)}_{002}  - N_2 B^{(2)}_{002} \bigg] \lb \tilde{\vec \ell}_1, \tilde{\vec \ell}_2, \tilde{\vec \ell}_3 \rb , \non\\
B^{\rm LO}_{{\rm n} s_2 s_3} \lb \vec \ell_1, \vec \ell_2, \vec \ell_3 \rb & =  \frac{1}{4} N_0 \int d\chi ~ \frac{W_{\rm n}(\chi) [W_g(\chi)]^2}{\chi^4}
\bigg[ N_0  B^{(0,0)}_{022} - N_2  B^{(\{0,2\})}_{022}   \non\\
&\hspace{4.5cm} - i \sqrt{2} N_1 \lb s_3 B^{(0,1)}_{022} + s_2 B^{(1,0)}_{022} \rb
\bigg]  \lb \tilde{\vec \ell}_1, \tilde{\vec \ell}_2, \tilde{\vec \ell}_3 \rb , \non\\
B^{\rm LO}_{s_1 s_2 s_3} \lb \vec \ell_1, \vec \ell_2, \vec \ell_3 \rb & =  \frac{1}{8} N_0^2 \int d\chi ~ \frac{[W_g(\chi)]^3}{\chi^4}
\bigg[
 N_0  B^{(0,0,0)}_{222}  - N_0^2 N_2 B^{(\{0,0,2\})}_{222} \non\\
&\hspace{4.5cm} - i \sqrt{2} N_0^2 N_1 \lb s_3 B^{(0,0,1)}_{222} + 2~{\rm cyc. } \rb \bigg]   \lb \tilde{\vec \ell}_1, \tilde{\vec \ell}_2, \tilde{\vec \ell}_3 \rb. \non
}
These results for $B^{\rm LO}_{{\rm n}{\rm n} s_3}$ agree with the projections used in Paper I \cite{Vlah++:2019} to decompose the bispectrum at tree-level in PT.

\bibliography{ms}

\end{document}